\documentclass[review]{elsarticle}

\usepackage{lineno,hyperref,amsmath,amssymb,mathtools,amsfonts,amsbsy,bm,geometry,multirow,epstopdf,graphicx,subcaption}
\modulolinenumbers[5]
\journal{Journal of Computational Physics}

\begin{document}

\begin{frontmatter}

\title{A target-fixed immersed-boundary formulation for rigid bodies interacting with fluid flow}

\author{Tzu-Yuan Lin}
\ead{d02522030@ntu.edu.tw}

\author{Hsin-Yu Hsieh}
\ead{r06522316@ntu.edu.tw}

\author{Hsieh-Chen Tsai\corref{corresp}}
\cortext[corresp]{Corresponding author}
\ead{hsiehchentsai@ntu.edu.tw}

\address{Department of Mechanical Engineering, National Taiwan University, Taipei, Taiwan, ROC}

\begin{abstract}
We present an immersed boundary projection method formulated in a body-fixed frame of reference for flow-structure interaction (FSI) problems involving rigid bodies with complex geometries. The body-fixed formulation is aimed at maximizing the accuracy of surface stresses on the FSI body (the target) on during spatial and temporal discretization. The incompressible vorticity equations and Newton's equations of motion are coupled implicitly so that the method remains stable for low solid-to-fluid mass ratios. The influence of fictitious fluid inside the rigid bodies is considered and the spurious oscillations in surface stresses are filtered to impose physically correct rigid body dynamics. Similar to many predecessors of the immersed boundary projection method, the resulting discrete system is solved efficiently using a block-LU decomposition. We then validate the method with two-dimensional test problems of a neutrally buoyant cylinder migrating in a planar Couette flow and a freely falling or rising cylindrical rigid body. 

\end{abstract}


\end{frontmatter}

%\linenumbers

\section{Introduction}
The immersed boundary (IB) method was originally proposed by Peskin \cite{peskin1972flow}. The key features of IB method are that the body surface is treated as a boundary immersed in the fluid and separate grids are used on the fluid and immersed boundary. Two grids communicate through smearing the surface stresses to the Eulerian fluid grid and interpolating the fluid velocity to the Lagrangian immersed boundary grid using a numerical delta function. This feature allows the method to handle objects with complex geometries interacting with fluid flow and obviate the need of computationally expensive tasks, such as re-meshing, since the Eulerian fluid grid does not need to conform with bodies. Taira and Colonius \cite{taira2007projection} developed a projection formulation of IB method which treats immersed boundary forces as Lagrange multipliers to the no-slip constraints to solve the incompressible flow around rigid bodies with prescribed kinematics. Colonius and Taira \cite{colonius2008fast} further improved the efficiency of the projection method by introducing a null-space approach of the method. The application of the immersed boundary projection method ranges from fundamental problems of airfoil aerodynamics \cite{taira2009tip,taira2009threedim,taira2010lift,chen2010,choi2015} and collisions between rigid bodies in fluids \cite{li2012contact}, to industrial problems in vertical-axis wind turbines \cite{tsai2016}.

During recent years, the IB method has been extensively developed and applied to flow-structure interaction problems for rigid bodies \cite{uhlmann2005,borazjani2008,eldredge2008,kempe2012,breugem2012,yang2015,wang2015,lacis2016stable} and elastic bodies \cite{zheng2010,borazjani2013fsi,goza2017coupled,tosi2019}. However, numerical instabilities due to added-mass effect of the fictitious fluid motion inside bodies when the motions of fluid and bodies are coupled explicitly (the coupling is so called weak coupling) has been reported by a number of studies \cite{uhlmann2005,borazjani2008,kempe2012,breugem2012}. The weak coupling algorithm becomes unstable when the solid-to-fluid mass ratio is reduced below some critical value. Borazjani {\it et al.} \cite{borazjani2008} showed that the numerical instability can be improved using an implicit coupling (a ``strong coupling"). Yang and Stern \cite{yang2015} presented a sharp-interface direct-forcing IB method that is strongly-coupled and non-iterative. Their method shows stable simulation for a low solid-to-fluid density ratio about 1. L{\=a}cis {\it et al.}\cite{lacis2016stable} discussed the stability properties between explicit and implicit coupling and proposed a strongly-coupled immersed boundary projection method that is stable for density ratios as low as $10^{-4}$. Alternatively, Eldredge \cite{eldredge2008} and Wang and Eldredge \cite{wang2015} have demonstrated that simulations at low density ratios can achieve better stability properties by including information of added mass in the numerical method.

Many IB methods are observed containing spurious oscillations in surface stresses \cite{yang2009,seo2011,goza2016accurate,kallemov2016}, which can result in inaccurate force and torque being exerted on bodies and lead to incorrect body dynamics for FSI problems. Goza {\it et al.} \cite{goza2016accurate} have noted that the surface stress oscillations are due to an inaccurate representation of the high frequency components of the surface stresses conveyed by the ill-posedness of an integral equation of the first kind by which surface stresses are solved. They also developed an efficient filtering technique to remove erroneous high-frequency stress components. 

In the present method,  we are interested in simulating rigid bodies that either are under flow-structure interaction (FSI bodies) or undergo prescribed motions in the fluid flow (non-FSI bodies). Particularly, we focus on obtaining the most accurate dynamics of an FSI body of interest (the target) by implementing three techniques. First, we formulate the general immersed-boundary governing equations in a frame of reference fixed on the target (see Appendix A for derivation of a simple form for the fluid equations in the body-fixed frame). The target-fixed nature of the formation allows the method to be free from iterating the position of the target; instead, we solve for the time integral of surface stresses between time steps. Second, the information of added mass of the fictitious fluid motion is included based on L{\=a}cis {\it et al.}\cite{lacis2016stable} for improving the numerical stability. Third, an accurate stress filter introduced by Goza {\it et al.} \cite{goza2016accurate} is used to impose to impose physically correct surface stresses. 

A null-space based fluid solver of Colonius and Taira \cite{colonius2008fast} is used to discretize the target-fixed fluid equations and coupled with the target-fixed equations of motion to form a discrete linear system of equations. The discrete equations is solved non-iteratively using a block-LU decomposition, which results in an algorithm with five steps. First, predictions are made for both fluid motion and FSI body kinematics in the absence of the surface stresses. Second, similar to L{\=a}cis {\it et al.}\cite{lacis2016stable}, a modified Poisson equation is solved for prediction of the surface stresses of FSI bodies that enforce the no-slip constraint and rigid body dynamics in the absence of non-FSI bodies. Third, another modified Poisson equation is solved for the surface stresses of non-FSI bodies that enforce the no-slip constraint. Fourth, the surface stresses of FSI bodies are corrected through projection. Finally, fluid motion and rigid body kinematics are updated through another projections. The current method is then validated with two tests: a circular cylinder freely falling or rising in fluids, and a neutrally buoyant cylinder migrating in a planar Couette flow.

\section{Governing equations}\label{sec_gov}
We consider rigid bodies immersed in an unbounded fluid domain $\Omega$ and the viscous flow induced by the rigid-body motion is incompressible. The FSI body is modeled by an immersed body $\Gamma_1$ and non-FSI bodies are modeled by immersed boundary bodies $\Gamma_2$. The system can be subject to a constant background acceleration such as the gravitational acceleration. The FSI body is free to translate and rotate about a center of rotation. The dynamics of the FSI body are under the influence of background acceleration and surface forces exerted by the fluid. Non-FSI bodies undergo prescribed motions in fluids. We have fixed the frame of reference on the FSI body (the target). In this body-fixed frame of reference, we let $\bm{x}$ denote the Eulerian coordinate representing a position vector in the fluid domain, $\bm{\chi}_1(s)$ as the Lagrangian coordinate attached to $\Gamma_1$, and $\bm{\chi}_2(s,t)$ as the Lagrangian coordinate attached to $\Gamma_2$, where $s$ is the variable that parameterizes the body surface. The dimensionless governing equations of this system are
\begin{align}
	\nonumber
	&\frac{\partial\bm{u}}{\partial t} = -\nabla \Pi\ + \left[\bm{u} - \bm{u}_a(\bm{x},t)\right]\times\bm{\omega} + \frac{1}{Re}\nabla^2\bm{u}\\ 
	&+ \int_{\Gamma_1} \bm{f}(\bm{\chi}_1(s),t)\delta(\bm{\chi}_1(s)-\bm{x})\ ds + \int_{\Gamma_2} \bm{f}(\bm{\chi}_2(s,t),t)\delta(\bm{\chi}_2(s,t)-\bm{x})\ ds\ ,\label{eq:momt_ib}\\
	&\nabla\cdot\bm{u} = 0\ ,\label{eq:incomp}\\
	&\int_\Omega\bm{u}(\bm{x},t)\delta(\bm{x}-\bm{\chi}_1(s))d\bm{x} = \bm{u}_s(t) + \bm{\omega}_s(t) \times \bm{r}(\bm{\chi}_1(s))\ ,\label{eq:no_slip1}\\
	&\int_\Omega\bm{u}(\bm{x},t)\delta(\bm{x}-\bm{\chi}_2(s,t))d\bm{x} = \bm{u}_B(\bm{\chi}_2(s,t),t)\ ,\label{eq:no_slip2}\\
	& M_e\left(\frac{d\bm{u}_s}{dt} + \bm{\omega}_s\times\bm{u}_s \right) =  - \int_{\Gamma_1} \bm{f}(\bm{\chi}_1(s),t)\ ds +  M_e\bm{a}\ , \label{eq_eom_trans}\\
	&\bm{I}_e\frac{d\bm{\omega}_s}{dt} + \bm{\omega}_s\times\left(\bm{I}_e\bm{\omega}_s\right) = - \int_{\Gamma_1} \bm{r}(\bm{\chi}_1(s)) \times \bm{f}(\bm{\chi}_1(s),t)\ ds +  M_e ( \bm{r}_c \times\bm{a})\ , \label{eq_eom_rot}
\end{align}
where 
\begin{align}
		&\Pi = p + \frac{1}{2}\left|\bm{u} - \bm{u}_a(\bm{x},t)\right|^2  - \frac{1}{2}\left|\bm{u}_a(\bm{x},t)\right|^2  +  \frac{1}{2}\left|\bm{u}_s\right|^2 - \bm{a}\cdot\bm{r}(\bm{x}), \label{eq_total_p}\\
		&\bm{u}_a(\bm{x},t) = \bm{u}_s(t) + \bm{\omega}_s(t) \times \bm{r}(\bm{x})\ ,\label{eq_ua}\\
		&M_e = M_s - V_s\ ,\label{eq_mass}\\
		&\bm{I}_e =  \bm{I}_s - \bm{I}_A\ .\label{eq_inertia}
\end{align}

In (\ref{eq:momt_ib}), a simple form of the fluid equation is derived in Appendix A. $\bm{u}$ and $\bm{\omega}$ are respectively the dimensionless fluid velocity and vorticity measured in the inertial frame. $\bm{f}$ are the dimensionless immersed boundary stresses exerted by the body on the fluid. $\Pi$ is the dimensionless modified pressure defined in (\ref{eq_total_p}). $p$ is the dimensionless pressure. $\bm{a}$ is the dimensionless time-varying background acceleration. $\bm{u}_a$ is the dimensionless velocity of the Eulerian fluid grid defined in (\ref{eq_ua}). $\bm{r}(\bm{x})$ is the dimensionless arm from the center of rotation to a point in the fluid domain.

(\ref{eq:incomp}) is the incompressibility constraint. (\ref{eq:no_slip1}) shows the no-slip condition for FSI body. The velocity on FSI body is composed of the translational and rotational components. $\bm{u}_s$ and $\bm{\omega}_s$ are the dimensionless center-of-mass translational and angular velocities and $\bm{r}(\bm{\chi}_1)$ is the dimensionless arm from the center of rotation to a point on $\Gamma_1$. (\ref{eq:no_slip2}) described the no-slip conditions for non-FSI bodies. The kinematics of non-FSI bodies are prescribed by the dimensionless velocity $\bm{u}_B(\bm{\chi}_2(s),t)$. 

 (\ref{eq_eom_trans}) is the translational equation of motion of the FSI body in the body-fixed frame. Based on L{\=a}cis {\it et al.}\cite{lacis2016stable}, the effect of the fictitious fluid motion inside the rigid body can be taken into account by introducing an effective mass, $M_e$, as in (\ref{eq_mass}). $V_s$, and $M_s$ are the dimensionless volume and mass of the FSI body. (\ref{eq_eom_rot}) is the rotational equation of motion of the FSI body in the body-fixed frame.  $\bm{r}_c$ is the dimensionless arm from the center of rotation to the center of mass of the rigid body. Similarly, an effective moment of inertia, $\bm{I}_e$, is introduced as in (\ref{eq_inertia}).  $\bm{I}_A$ and $\bm{I}_s$ are the dimensionless area moment of inertia and moment of inertia of the body about the center of rotation. We consider the axes of the body-fixed frame to be aligned with the principal axes of $\bm{I}_s$, so that $\bm{I}_s$ is diagonal. If a thin rigid body is considered, then $V_s$ and $\bm{I}_A$ are equal to zero. 

In (\ref{eq:momt_ib}) - (\ref{eq_inertia}), $\bm{r}$, $\bm{x}$, $\bm{\chi}_i$, and $s$ were nondimensionalized by a characteristic length scale, $L$; $\bm{u}$, $\bm{u}_s$, and $\bm{u}_B$ were nondimensionalized by a characteristic velocity scale, $U_\infty$; $\nabla$, $V_s$, and $\bm{I}_A$ were nondimensionalized by $1/L$, $L^3$, and $L^5$; $t$ was nondimensionalized by $L/U_\infty$; $\bm{\omega}$ and $\bm{\omega}_s$ were nondimensionalized by $U_\infty/L$; $\bm{a}$ was nondimensionalized by $U_\infty^2/L$; $M_s$ was nondimensionalized by $\rho_f L^3$;  $\bm{I}_s$ was nondimensionalized by $\rho_f L^5$; and $p$ and $\bm{f}$ were nondimensionalized by $\rho_fU_\infty^2$, where $\rho_f$ is the fluid density. The Reynolds number in (\ref{eq:momt_ib}) is defined as $Re=U_\infty L/\nu_f$, where $\nu_f$ is the kinematic viscosity of the fluid.

The advantage of taking a frame of reference fixed on the body of interest is that the immersed boundary does not move in this frame of reference, so that the spatial integrations over $\Gamma_1$ in (\ref{eq:momt_ib}), (\ref{eq_eom_trans}), and (\ref{eq_eom_rot}) can be pull out of time integral at temporal discretization. The price for this simplicity is that we must now use the Navier-Stokes equations in a body-fixed frame (See (\ref{eq:non-inertial}) in Appendix A) and correct time derivative of $M_e\bm{u}_s$ and $\bm{I}_e\bm{\omega}_s$ with their cross products with body angular velocity in (\ref{eq_eom_trans}) and (\ref{eq_eom_rot}). Fortunately, a simple form of (\ref{eq:non-inertial}) derived in Appendix A and corrections to $M_e\bm{u}_s$ and $\bm{I}_e\bm{\omega}_s$ can be easily implemented. 

For convenience, we can write the cross product of two vectors $\bm{a}$ and $\bm{b}$ in the following matrix operation:
\begin{align}
	\bm{a}\times\bm{b} = \bm{X}(\bm{a})\bm{b} = \bm{X}^T(\bm{b})\bm{a}\ ,
\end{align}
where
\begin{align}
	\bm{X}(\bm{a}) = \begin{pmatrix}
	0 & -a_3 & a_2\\
	a_3 & 0 & -a_1\\
	-a_2 & a_1 & 0
	\end{pmatrix}
\end{align}
is a skew-symmetric matrix associated with the vector $\bm{a} = (a_1,a_2,a_3)^T$. Therefore, (\ref{eq:momt_ib}), (\ref{eq:no_slip1}), (\ref{eq_eom_trans}), (\ref{eq_eom_rot}), and (\ref{eq_ua}) can be written as 
\begin{align}
	\nonumber
	&\frac{\partial\bm{u}}{\partial t} = -\nabla \Pi\ + \bm{X}\left(\bm{u} - \bm{u}_a(\bm{x},t)\right)\bm{\omega} + \frac{1}{Re}\nabla^2\bm{u}\\
	&+ \int_{\Gamma_1} \bm{f}(\bm{\chi}_1(s),t)\delta(\bm{\chi}_1(s)-\bm{x})\ ds + \int_{\Gamma_2} \bm{f}(\bm{\chi}_2(s,t),t)\delta(\bm{\chi}_2(s,t)-\bm{x})\ ds\ ,\label{eq:momt_ib2}\\
	&\int_\Omega\bm{u}(\bm{x},t)\delta(\bm{x}-\bm{\chi}_1(s))d\bm{x} = \bm{u}_s(t) + \bm{X}^T(\bm{r}(\bm{\chi}_1(s))) \bm{\omega}_s(t)\ ,\label{eq:no_slip1_2}\\
	& M_e\left(\frac{d\bm{u}_s}{dt} + \bm{X}(\bm{\omega}_s)\bm{u}_s \right) =  - \int_{\Gamma_1} \bm{f}(\bm{\chi}_1(s),t)\ ds +  M_e\bm{a}\ , \label{eq_eom_trans2}\\
	&\bm{I}_e\frac{d\bm{\omega}_s}{dt} + \bm{X}(\bm{\omega}_s)\bm{I}_e\bm{\omega}_s = - \int_{\Gamma_1} \bm{X}(\bm{r}(\bm{\chi}_1(s))) \bm{f}(\bm{\chi}_1(s),t)\ ds +  M_e \bm{X}(\bm{r}_c)\bm{a}\ , \label{eq_eom_rot2}\\
	&\bm{u}_a(\bm{x},t) = \bm{u}_s(t) + \bm{X}^T(\bm{r}(\bm{x}))\bm{\omega}_s(t)\ .\label{eq_ua2}
\end{align}

\section{Numerical method}\label{sec_num}

In this section, we first discretize the governing equations in space to obtain time-dependent semi-discrete equations. The fluid equations are discretized using the 2D discrete streamfunction formulation developed by Colonius and Taira \cite{colonius2008fast}. The semi-discrete equations are further discretized in time and coupled with the equations of motion integrated in time. The fully discrete algebraic equations are solved by a projection technique associated with the block-LU decomposition.

\subsection{Spatial discretization and accurate stress filter}
	Following Taira and Colonius \cite{taira2007projection}, we consider the spatial discretization in the fluid domain on a two-dimensional unbounded uniform staggered Cartesian grid, and the spatial discretization on the immersed boundary on an evenly spaced grid. Moreover, all the grid spacings are set to be the same, i.e., $\Delta x = \Delta y = \Delta s$. We start with discretizing the equations of motion of the rigid body spatially as
\begin{align}
	&M_e\left(\frac{d\bm{u}_s}{dt} + \bm{X}(\bm{\omega}_s)\bm{u}_s \right) = S_1W_1 f_1 + M_e\bm{a}\ , \label{eq_eom_trans_semi}\\
	&\bm{I}_e\frac{d\bm{\omega}_s}{dt} + \bm{X}(\bm{\omega}_s)\bm{I}_e\bm{\omega}_s  = X_1 W_1 f_1 +  M_e \bm{X}(\bm{r}_c)\bm{a}\ , \label{eq_eom_rot_semi}
\end{align}
where $f_1$ are the spatially discrete surface stresses of the FSI body (the spatial discretization of $-\bm{f}(\bm{\chi}_1(s),t)$). 

Since the surface stresses obtained by many immersed boundary methods contain spurious oscillations \cite{yang2009,seo2011,goza2016accurate}, we include an accurate stress filter, $W_1$, introduced by Goza {\it et al.} \cite{goza2016accurate} in (\ref{eq_eom_trans_semi}) and (\ref{eq_eom_rot_semi}) to obtain the physically correct surface stress on the immersed boundary. $W_1f_1$ is a discretization of 
\begin{align}
	\frac{\int_\Omega\int_{\Gamma_1}(-\bm{f}(\bm{\chi}_1(s)))\delta(\bm{x}-\bm{\chi}_1(s))\delta(\bm{\chi}_1(s)-\bm{x})\ ds \ d\bm{x}}{\int_{\Gamma_1}\delta(\bm{x}-\bm{\chi}_1(s))\ ds}\ .
\end{align}
The numerical delta function used in the present work is from Roma \cite{roma1999}. The specific form of $W_1$ can be found in reference \cite{goza2016accurate}. We also note that $W$ is a diagonal matrix so that $W_1^T = W_1$. Moreover, the linear integration operator, $S_1$, is the spatial discretization of $\int_{\Gamma_1}(\cdot)\ ds$. The angular integration operator, $X_1$, is the spatial discretization of $\int_{\Gamma_1} \bm{X}(\bm{r}(\bm{\chi}_1(s))) (\cdot)\ ds$.

By defining $M = diag(M_e\bm{I}_3, \bm{I}_e)$ ($\bm{I}_3$ is a $3\times3$ identity matrix), $\lambda = (\bm{u}_s,\bm{\omega}_s)^T$, $Q = (S_1 W_1, X_1 W_1)^T$ and $R = (M_e\bm{a} - M_e\bm{X}(\bm{\omega}_s)\bm{u}_s, M_e\bm{X}(\bm{r}_c)\bm{a} - \bm{X}(\bm{\omega}_s)\bm{I}_e\bm{\omega}_s)^T$, (\ref{eq_eom_trans_semi}) and (\ref{eq_eom_rot_semi}) can be written as
\begin{align}
	M\frac{d\lambda}{dt} = Qf + R\ .\label{eq_eom_semi}
\end{align}

Next, we discretize the fluid equations spatially as
\begin{align}
	\frac{dq}{dt} &= -G {\it \Pi} + N(q,q_a) + \frac{1}{Re} L q - H_1 f_1 - H_2 f_2 \ ,\label{eq:momt_ib_semi}\\
	D q &= 0 \ ,\label{eq:cont_semi}\\
	E_1 q &= W_1^TS_1^T \bm{u}_s + W_1^TX_1^T\bm{\omega}_s = Q^T\lambda\ ,\label{eq:no_slip1_filtered}\\
	E_2 q &= q_{B_2}\ ,\label{eq:no_slip2_discrete}
\end{align} 
where $q$ and ${\it \Pi}$ are the spatially discrete fluid flux and pressure. $q_{B_2}$ is the spatial discretization of $\bm{u}_B(\bm{\chi}_2(s,t),t)$. $G$, $D$, and $L$ are the discrete gradient, divergence, and Laplacian operators. $N(q,q_a)$ is the spatial discretization of the nonlinear term $\bm{X}\left(\bm{u} - \bm{u}_a(\bm{x},t)\right)\bm{\omega}$. $H_i$ and $E_i$ are the discretizations of the regularization and interpolation operators with respect to $\Gamma_i$ in (\ref{eq:momt_ib}), (\ref{eq:no_slip1}), and (\ref{eq:no_slip2}). Discrete operators and variables are scaled such that $D = - G^T$ and $H_i = E_i^T$ for computational efficiency. We note that $W_1^T$ is included in both terms on the right-hand side of (\ref{eq:no_slip1_filtered}) in order to write the equations in a symmetric form later. Since both $S_1^T \bm{u}_s$ and $X_1^T\bm{\omega}_s$ have no spurious oscillations, $W_1^T$ has no effect on the resulting fluxes.

	Following Colonius and Taira \cite{colonius2008fast}, a discrete curl operator, $C$, which lies in the null space of the discrete divergence operator, $D$, is constructed to mimic the vector identities that the divergence of the curl of any vector field and the curl of the gradient of any scalar field are always zero, i.e., $DC = -(C^TG)^T = 0$. By introducing the discrete streamfunction, $s$, such that $q = Cs$ and taking curl, $C^T$, of (\ref{eq:momt_ib_semi}), the incompressibility constraint (\ref{eq:cont_semi}) is satisfied automatically and the $-G{\it \Pi}$ term in  (\ref{eq:momt_ib_semi}) can be dropped. The fluid equations can be written as
\begin{align}
	C^TC\frac{ds}{dt} &= C^TN(q,q_a(x)) + \frac{1}{Re} C^TLCs - C^TE_1^T f_1 -  C^TE_2^T f_2 \ ,\label{eq:momt_ib_semi2}\\
	E_1Cs &= Q^T\lambda\ ,\label{eq:no_slip1_sf}\\
	E_2Cs &= q_{B_2}\ .\label{eq:no_slip2_sf}
\end{align} 
(\ref{eq:momt_ib_semi2}) is nothing but the semi-discrete vorticity equation with immersed-boundary forcing in the body-fixed frame and can be easily modified from the original null-space-approach immersed boundary projection method developed by Colonius and Taira \cite{colonius2008fast}.

\subsection{Temporal discretization and factorization procedure}
	For the temporal discretization of the governing equations, we integrate (\ref{eq_eom_semi}) and (\ref{eq:momt_ib_semi2}) numerically from time $t_n$ to $t_{n+1}$. We use an Adams-Bashforth scheme for the $R$ term in  (\ref{eq_eom_semi}) and the nonlinear terms in (\ref{eq:momt_ib_semi2}), a Crank-Nicolson method for the diffusive term in (\ref{eq:momt_ib_semi2}), and an implicit Euler for $C^TE_2^Tf_2$ in (\ref{eq:momt_ib_semi2}). Most importantly, since $E_1$ and $W_1$ are time-invariant in the body-fixed frame and $C^T$, $S_1$, and $X_1$ are independent of time, we can pull them out of the time integrals and integrate $f_1$ directly.  We define the average immersed boundary surface stress $\overline{f}_1$ in $[t_n,t_{n+1}]$, i.e., $\overline{f}_1 = \frac{1}{\Delta t}\int_{t_n}^{t_{n+1}} f_1 dt$. Moreover, we evaluate (\ref{eq:no_slip1_sf}) and (\ref{eq:no_slip2_sf}) at $t_{n+1}$. After the temporal discretization, the semi-discrete governing equations yield a linear system:
\begin{align}
	\begin{bmatrix}
		C^TAC & 0 & C^TE_1^T & C^TE_2^{{n+1}^T}\\
		0 & M & -Q & 0\\
		E_1C & -Q^T & 0 & 0\\
		E_2^{n+1}C & 0 & 0 & 0
	\end{bmatrix}
	\begin{pmatrix}
		s^{n+1}\\
		\lambda^{n+1}\\
		\overline{f}_1 \Delta t\\
		f_2^{n+1}\Delta t
	\end{pmatrix}
	=
	\begin{pmatrix}
		r_1^n\\
		r_2^n\\
		0\\
		q_{B_2}^{n+1}
	\end{pmatrix}\ , \label{eq:disc_sys}
\end{align}
where 
\begin{align}
	A &= I -\frac{\Delta t}{2Re}L\ ,\\
	r_1^n &= C^T\left(I +\frac{\Delta t}{2Re}L\right)Cs^n + \frac{3\Delta t}{2}C^TN(q^n,q_a^n) - \frac{\Delta t}{2}C^TN(q^{n-1},q^{n-1}_a)\ ,\\
	r_2^n &= M\lambda^n + \frac{3\Delta t}{2}R^n - \frac{\Delta t}{2}R^{n-1}\ .
\end{align}

We use the block-LU decomposition to factorize the linear system (\ref{eq:disc_sys}). The factored equations are given below:
\begin{align}
	&\begin{bmatrix}
		C^TAC & 0\\
		0 & M 
	\end{bmatrix}
	\begin{pmatrix}
		s^*\\
		\lambda^*
	\end{pmatrix}
	=
	\begin{pmatrix}
		r_1^n\\
		r_2^n
	\end{pmatrix}\ , \label{eq_pred}\\
	[Q^TM^{-1}&Q + P_{11}]\overline{f}_1^*\Delta t = E_1Cs^*-Q^T\lambda^* \ , \label{eq_poisson1}\\
	\left[P_{22}^{n+1} - P_{21}^{n+1}(Q^TM^{-1}Q + P_{11})^{-1}\right.&\left.P_{12}^{n+1}\right]f_2^{n+1}\Delta t = E_2^{n+1}Cs^*- q_{B_2}^{n+1}-P_{21}^{n+1}\overline{f}_1^*\Delta t\ , \label{eq_poisson2}\\
	\overline{f}_1\Delta t = \overline{f}_1^*\Delta t - &[Q^TM^{-1}Q + P_{11}]^{-1}P_{12}^{n+1}f_2^{n+1}\Delta t\ , \label{eq_projf}\\
	 \begin{pmatrix}
		s^{n+1}\\
		\lambda^{n+1}
	\end{pmatrix} 
	=
	\begin{pmatrix}
		s^*\\
		\lambda^*
	\end{pmatrix} + &\begin{pmatrix}
		-(C^TAC)^{-1}C^T\left(E_1^T\overline{f}_1\Delta t+E_2^{{n+1}^T}f_2^{n+1}\Delta t\right)\\
		M^{-1}Q\overline{f}_1\Delta t 
	\end{pmatrix}\ , \label{eq_proj}
\end{align}
where
\begin{align}
	P_{11} &=  E_1C(C^TAC)^{-1}C^TE_1^T\ ,\\
	P_{12}^{n+1} &=  E_1C(C^TAC)^{-1}C^TE_2^{{n+1}^T}\ ,\\
	P_{21}^{n+1} &=  E_2^{n+1}C(C^TAC)^{-1}C^TE_1^T\ ,\\
	P_{22}^{n+1} &=  E_2^{n+1}C(C^TAC)^{-1}C^TE_2^{{n+1}^T}\ .
\end{align}

The factored equations (\ref{eq_pred})-(\ref{eq_proj}) are analogous to the fractional-step procedure for the Navier-Stokes equations \cite{perot1993}. Analogous fractional-step methods have been developed by Taira and Colonius \cite{taira2007projection} for rigid body undergoing prescribed motions, L{\=a}cis {\it et al.}\cite{lacis2016stable} for rigid-body interacting with the flow, and Goza and Colonius \cite{goza2017coupled} for flow-structure interaction of thin elastic structures. Our current method is approximate to Taira and Colonius \cite{taira2007projection} when there are only non-FSI bodies and L{\=a}cis {\it et al.}\cite{lacis2016stable} when there is only FSI body, differing in that all variables in the target-fixed frame and the surface stresses of FSI body have been physically-corrected and time-averaged between each time step. 
 
The physical interpretation of (\ref{eq_pred}) is that a trial streamfunction and trial rigid-body kinematics are predicted by evolving the discrete fluid equations and the equations of motion in the absence of the immersed boundary forcing. In (\ref{eq_poisson1}), a Poisson-like problem for the prediction of the surface stresses of FSI body is solved to enforce the no-slip condition and the rigid-body dynamics in the absence of non-FSI bodies. In (\ref{eq_poisson2}), another Poisson-like problem for the surface stresses of non-FSI bodies is solved to enforce the no-slip condition. The influence of the prediction of the surface stresses of FSI body to non-FSI bodies has also been taken into account. Through the projection step shown in (\ref{eq_projf}), the surface stresses of FSI body is updated to include the influence of the surface stresses of non-FSI bodies. Finally, in (\ref{eq_proj})  the streamfunction and rigid-body kinematics are corrected through projections to remove the part of the trial solution that does not satisfy the constraints.

When we solve the trial streamfunction in (\ref{eq_pred}), a multi-domain setting is used to account for the boundary condition at infinity (see details in reference \cite{colonius2008fast}) and the Poisson-like problem $(C^TAC)^{-1}$ can be solved efficiently using the discrete sine transform.  In (\ref{eq_poisson1}), since $Q^TM^{-1}Q + P_{11}$ is time-invariant and symmetric, it can be precomputed and solved efficiently (using, for example, the Cholesky decomposition).  In (\ref{eq_poisson2}), $P_{22}^{n+1} - P_{21}^{n+1}(Q^TM^{-1}Q + P_{11})^{-1}P_{12}^{n+1}$ is symmetric and can be solved efficiently using an iterative method such as Conjugate Gradient method. Once $\lambda^{n+1}$ is computed, the location of the center of mass and the rotation angle of the rigid body can be integrated numerically as a post-processing step using an explicit scheme, like an Adams-Bashforth scheme.

\section{Verifications and validations}\label{sec_validate}
\subsection{A neutrally buoyant cylinder migrating in a planar Couette flow}
\begin{figure}[h!]
	\centering
	\begin{subfigure}[b]{0.6\textwidth}
	\centering
		\includegraphics[width=\textwidth]{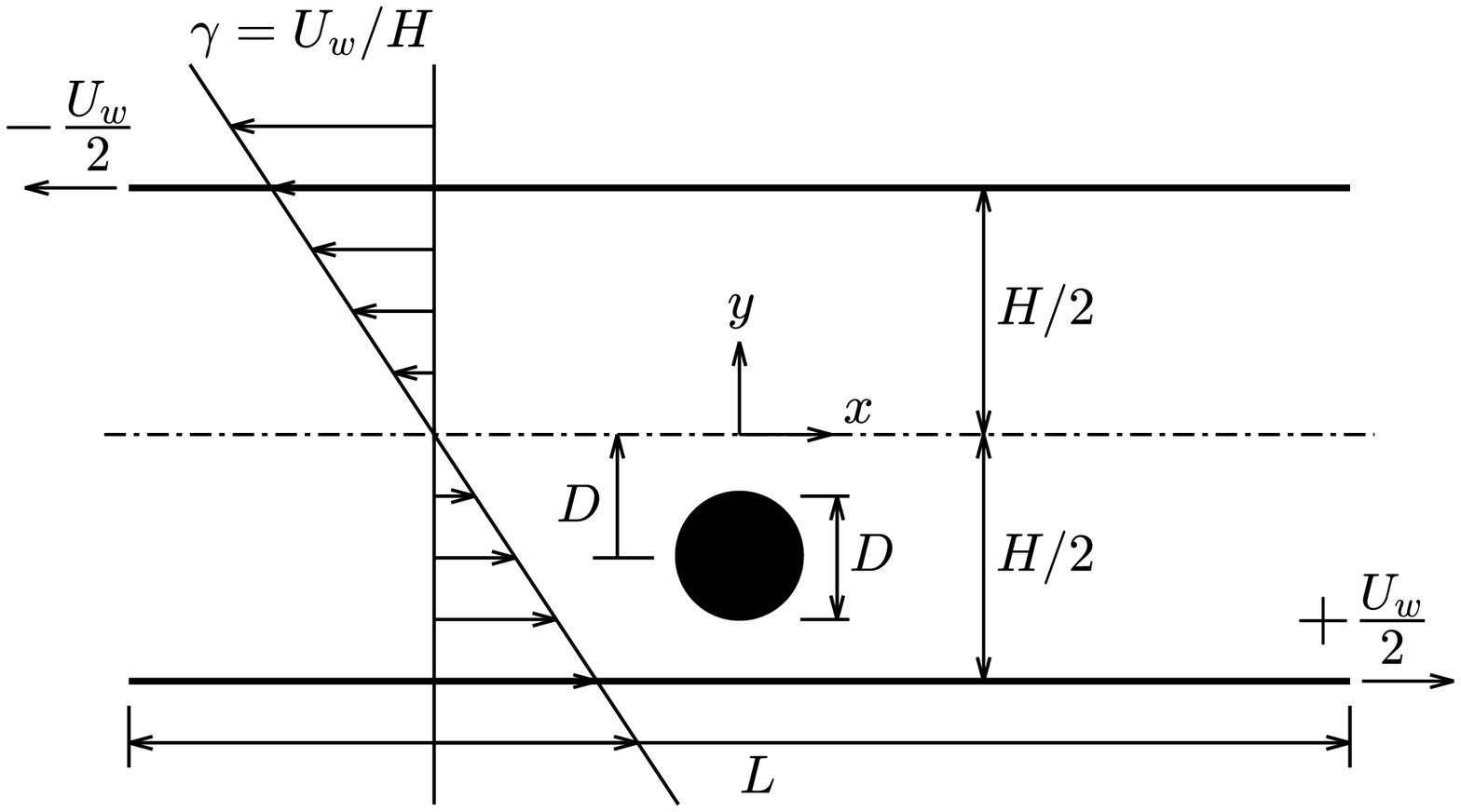}
		\caption{ }
		\label{fig_schematic_couette}
	\end{subfigure}
	\begin{subfigure}[b]{0.45\textwidth}
	\centering
		\includegraphics[width=0.58\textwidth]{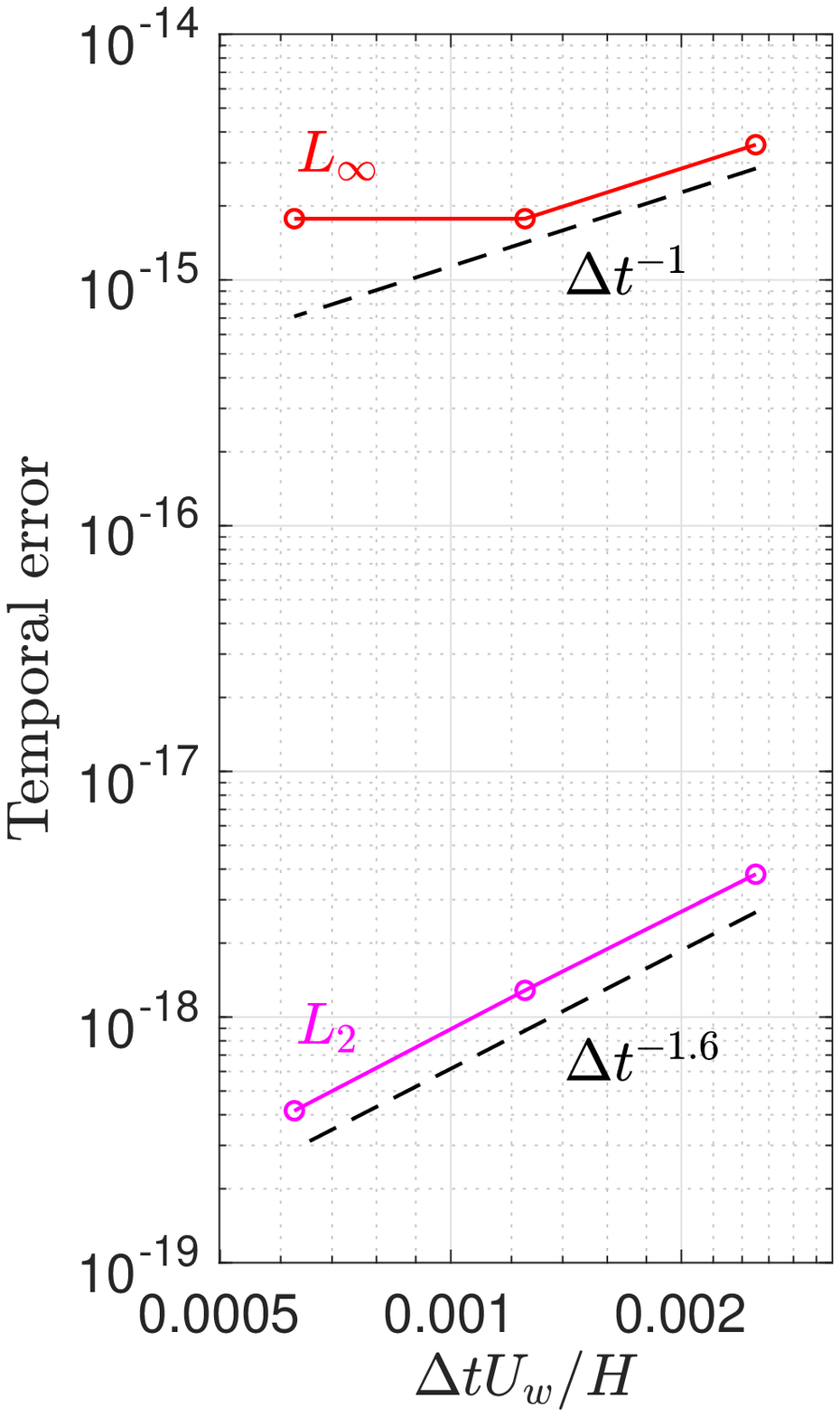}
		\caption{ }
		\label{fig_temporal_convergence}
	\end{subfigure}
	\begin{subfigure}[b]{0.45\textwidth}
	\centering
		\includegraphics[width=0.6\textwidth]{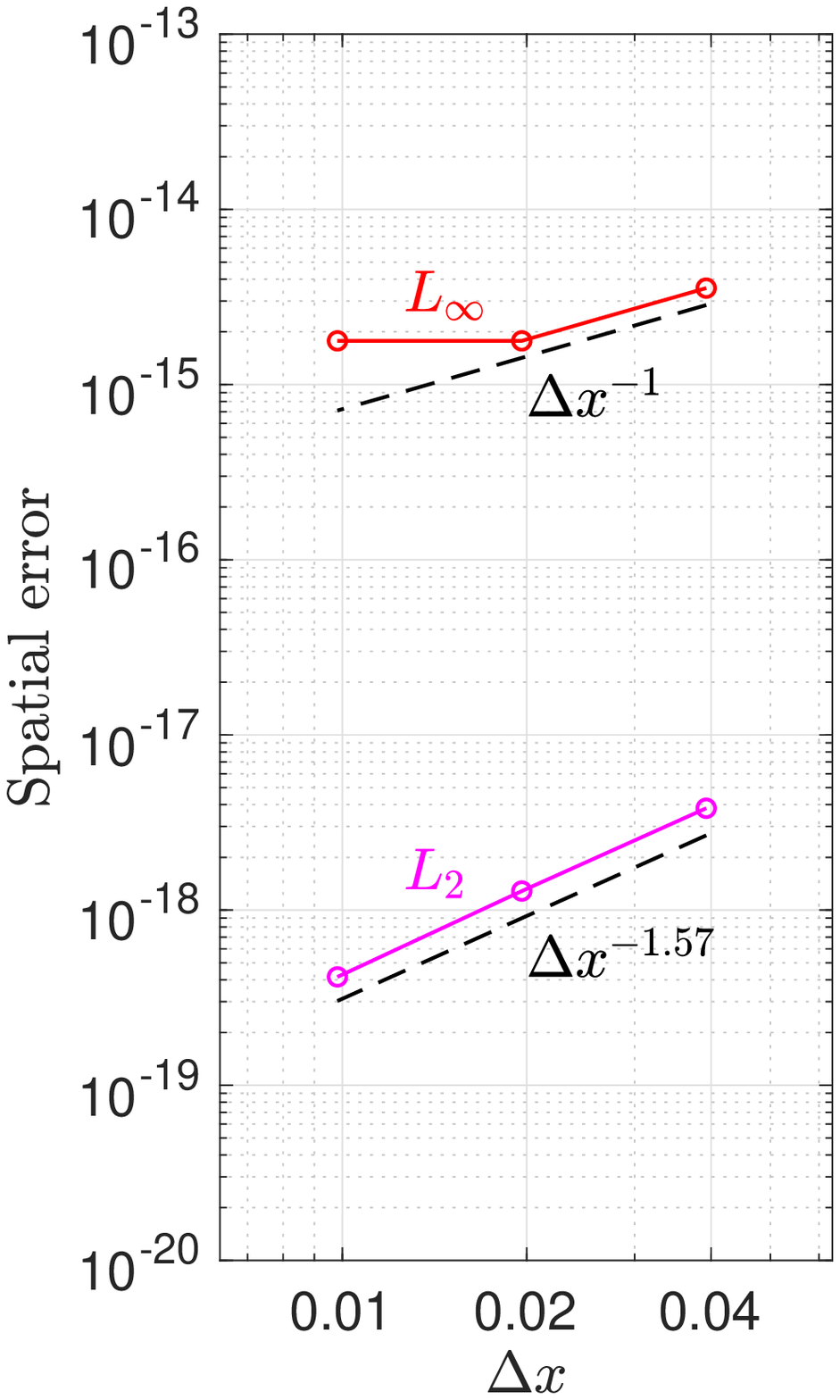}
		\caption{ }
		\label{fig_spatial_convergence}
	\end{subfigure}
	\caption{(a) The schematic of a neutrally buoyant cylinder migrating in a planar Couette flow and (b) the temporal and (c) spatial convergence of the present method.}
	\label{fig_couette_flow}
\end{figure}

In order to validate the present method, we consider a test problem of a neutrally buoyant cylinder migrating in a planar Couette flow.  As shown in figure \ref{fig_schematic_couette}, a neutrally buoyant cylinder with radius $D$ is initially at rest and released at $y=-D$. The cylinder is free to translate and rotate. In order to mimic a planar Couette flow in a channel of height $H$, two horizontal flat plates with length $L$ are placed at $y=\pm H/2$ in a simple shear flow. The channel height $H=4.0D$ and the plate length is $L=15.0D$. The shear rate of the simple shear flow is $\gamma = U_w/H$ and the upper and lower plates are moving in $x$-direction with velocities $-U_w/2$ and $U_w/2$, respectively. This configuration is also used by L{\=a}cis {\it et al.}\cite{lacis2016stable}, Feng {\it et al.}\cite{feng1994}, and Vasseur and Cox\cite{vasseur1976}.

In the following simulations, the characteristic length is the channel height $H$ and the characteristic velocity is $U_w$. Fluid simulations are done in a frame of reference traveling together with the cylinder. We use a multi-domain setting of the first domain size $4H\times2H$ and the number of domains $N_g = 4$.  As shown in (\ref{eq_mass}) and (\ref{eq_inertia}), $M_e$ and $\bm{I}_e$ is singular when $\rho = \rho_s/\rho_f = 1$, where $\rho_s$ is the solid density. Therefore, we use a slightly larger density ratio $\rho = 1.01$ and set the background acceleration $\bm{a} = 0$. In order to compare the numerical solution with the works by L{\=a}cis {\it et al.}\cite{lacis2016stable} and Feng {\it et al.}\cite{feng1994}, a Reynolds number $Re_H = U_w H/\nu_f = 40$ is selected. Feng {\it et al.}\cite{feng1994} used a finite element solver on a body-fitted mesh and L{\=a}cis {\it et al.}\cite{lacis2016stable} used an implicitly-coupled immersed boundary projection method in a finite domain of size $40H\times H$ with velocity Dirichlet boundary conditions being specified. 

First we focus on the convergence properties of the present method when dynamics of both FSI and non-FSI bodies are coupled with the fluid solver. In order to investigate the temporal convergence, four sets of $(\Delta x/H,\Delta t U_w/H)$ are used: $(10^{-2},2.5\times10^{-3})$, $(5\times10^{-3},1.25\times10^{-3})$, $(2.5\times10^{-3},6.25\times10^{-4})$, and $(1.25\times10^{-3},3.125\times10^{-4})$, which give CFL numbers around 0.25. The simulations are carried out until $t U_w/H=0.25$ and the simulation with the smallest $\Delta t$ is used as reference to compute errors in the vorticity field. The temporal convergences of the infinity norm $L_\infty$ and the 2 norm $L_2$ of errors are shown in figure \ref{fig_temporal_convergence}. We can see that the convergence rate in time is around 1 in $L_\infty$ and around 1.6 in $L_2$. To investigate the spatial convergence, we use a very small time step of $\Delta t U_w/H = 10^{-4}$ and grid spacing of $\Delta x/H =$ 0.01, 0.005, 0.0025, and 0.00125. We carry out the simulations until $t U_w/H=0.01$ and the simulation with the finest grid is used as reference to compute the vorticity errors. As shown in figure \ref{fig_spatial_convergence}, the convergence rate in space is around 1 in $L_\infty$ and around 1.57 in $L_2$.

\begin{figure}[h!]
	\centering
	\begin{subfigure}[b]{0.3\textwidth}
		\includegraphics[width=\textwidth]{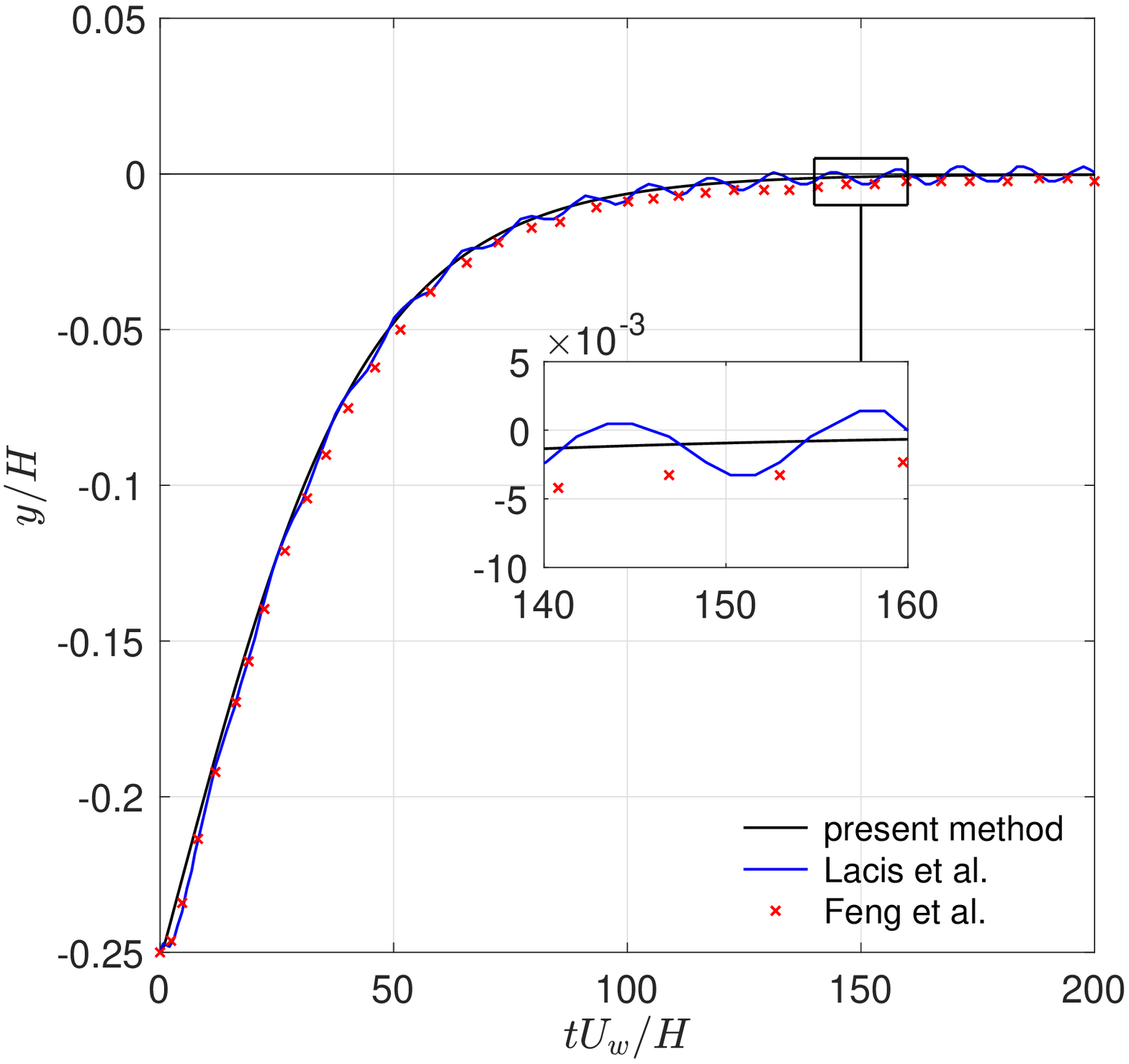}
		\caption{ }
		\label{fig_ycmL}
	\end{subfigure}
	\begin{subfigure}[b]{0.3\textwidth}
		\includegraphics[width=\textwidth]{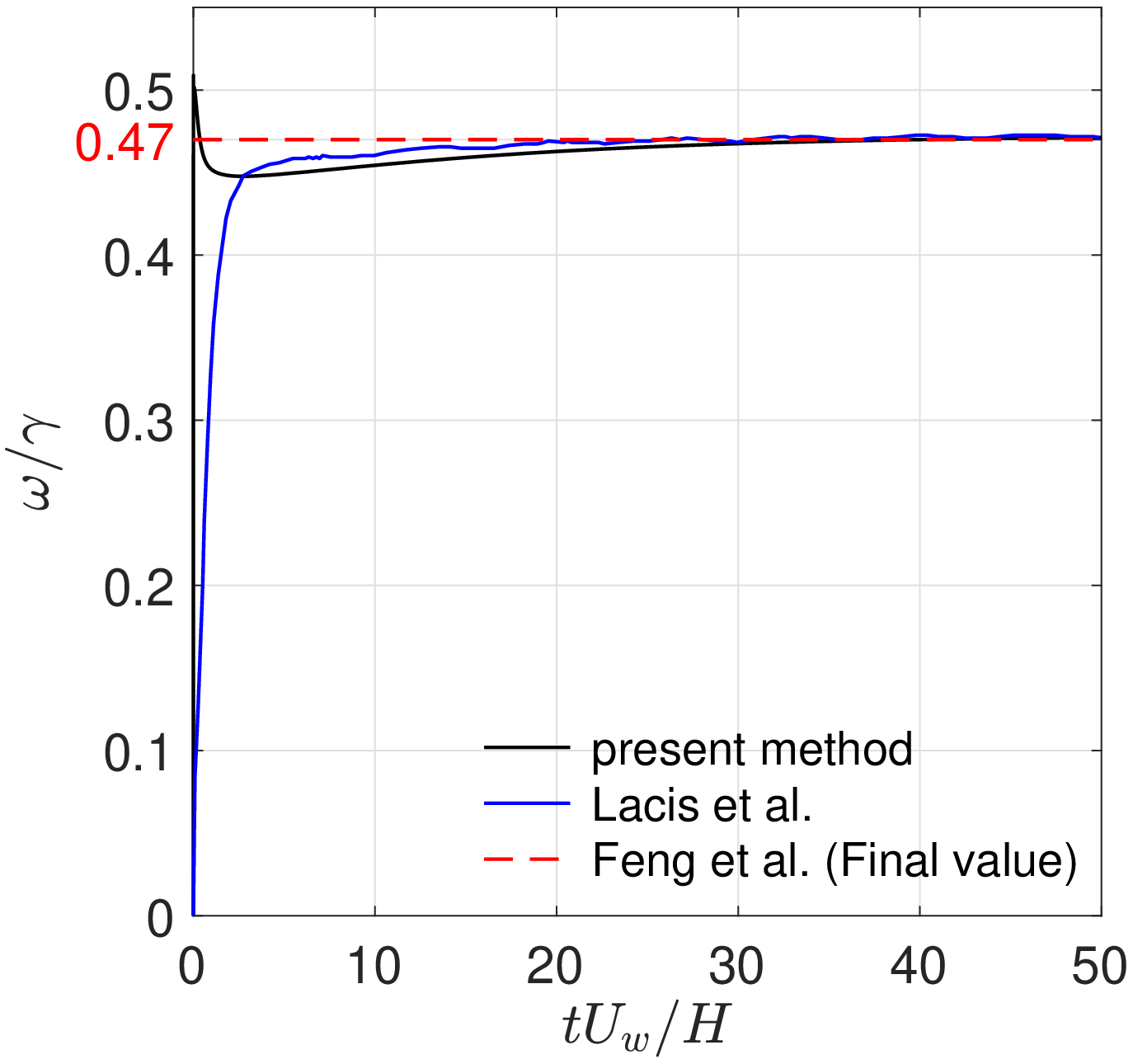}
		\caption{ }
		\label{fig_omega}
	\end{subfigure}
	\begin{subfigure}[b]{0.315\textwidth}
		\includegraphics[width=\textwidth]{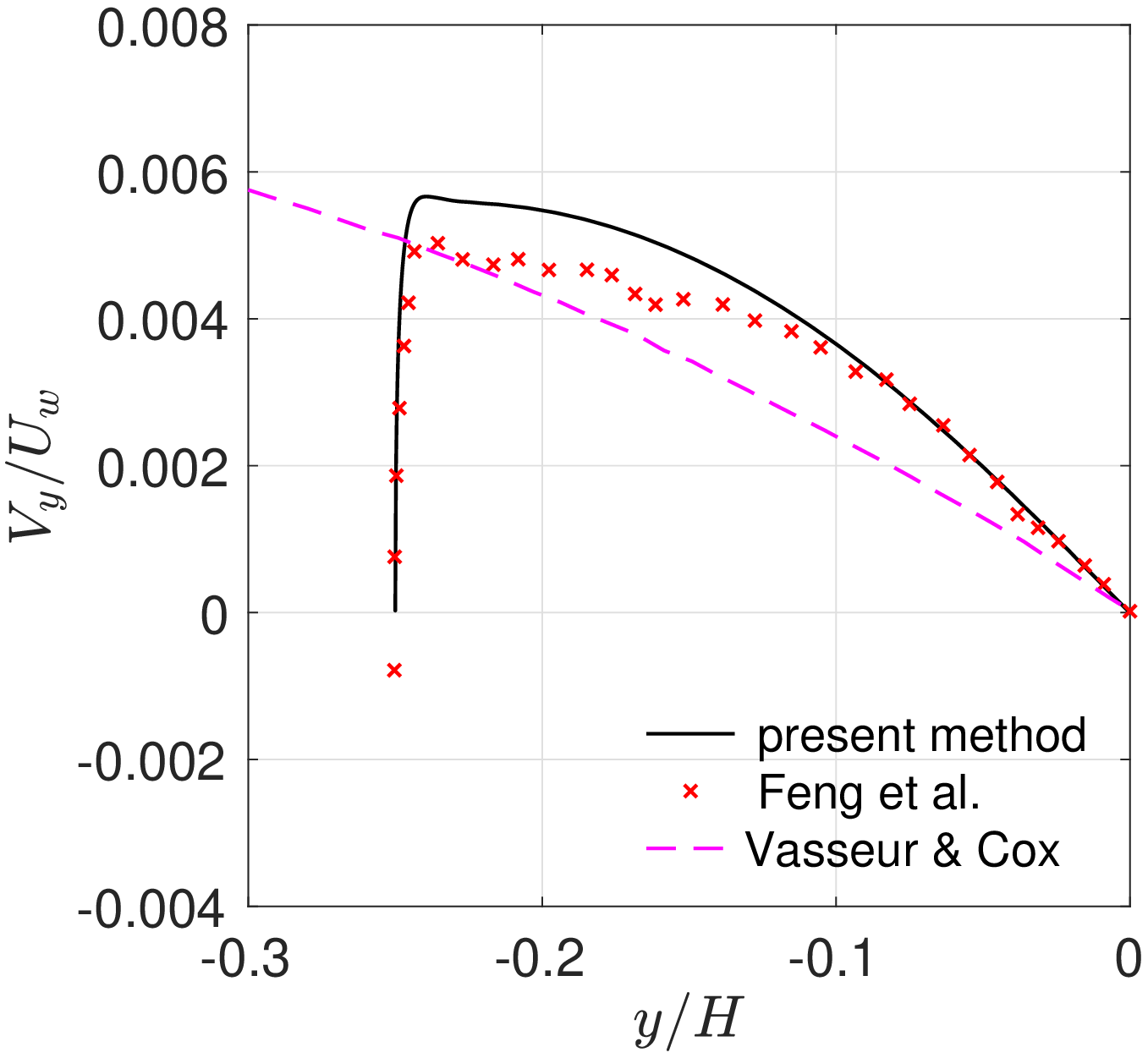}
		\caption{ }
		\label{fig_vcmL}
	\end{subfigure}
	\caption{The histories of (a) the vertical displacement and (b) the angular velocity of the neutrally buoyant cylinder, and (c) the phase space of the vertical displacement of the neutrally buoyant cylinder. Only the final value of the ratio of the angular velocity to the shear rate was reported by Feng {\it et al.}\cite{feng1994}.}
	\label{fig_migration}
\end{figure}

In the following simulation we use a moderate grid spacing $\Delta x = 0.01 H$ and a moderate time step $\Delta t = 0.005 H/U_w$. Figures \ref{fig_ycmL} and \ref{fig_omega} show the histories of the vertical displacement and the angular velocity of the neutrally buoyant cylinder, respectively. The results agree well with L{\=a}cis {\it et al.}\cite{lacis2016stable} and Feng {\it et al.}\cite{feng1994}. The cylinder is initially at rest. Once the cylinder is released, it migrates downstream and toward the center of the channel and its rotating rate increase rapidly to a final value about 47\% of the shear rate of the simple shear flow. According to L{\=a}cis {\it et al.}\cite{lacis2016stable}, the small oscillations in their trajectory (figure \ref{fig_ycmL}) are due to the fact that the Lagrangian points are moving relative to the fluid grid during the migration. They referred this oscillating behavior as ``grid locking”, where similar behavior was also observed in IB simulations by Breugem \cite{breugem2012}. Due to the target-fixed nature of the present method, we avoid generating the ``grid-locking” oscillations. Figure \ref{fig_vcmL} shows the vertical displacement and the velocity velocity of the neutrally buoyant cylinder in the phase space. The numerical result of Feng {\it et al.} \cite{feng1994} and the analytical result of Vasseur and Cox \cite{vasseur1976} for a small sphere migrating in a slow flow are also shown for comparison. Besides the initial transients, which are not being addressed by Vasseur and Cox \cite{vasseur1976}, the trends are alike and the computational results agree well when the cylinder moves closer to the center of the channel.

\subsection{Freely falling/rising circular cylinders}
\begin{figure}[h!]
	\centering
	\begin{subfigure}[b]{0.30\textwidth}
		\includegraphics[width=\textwidth]{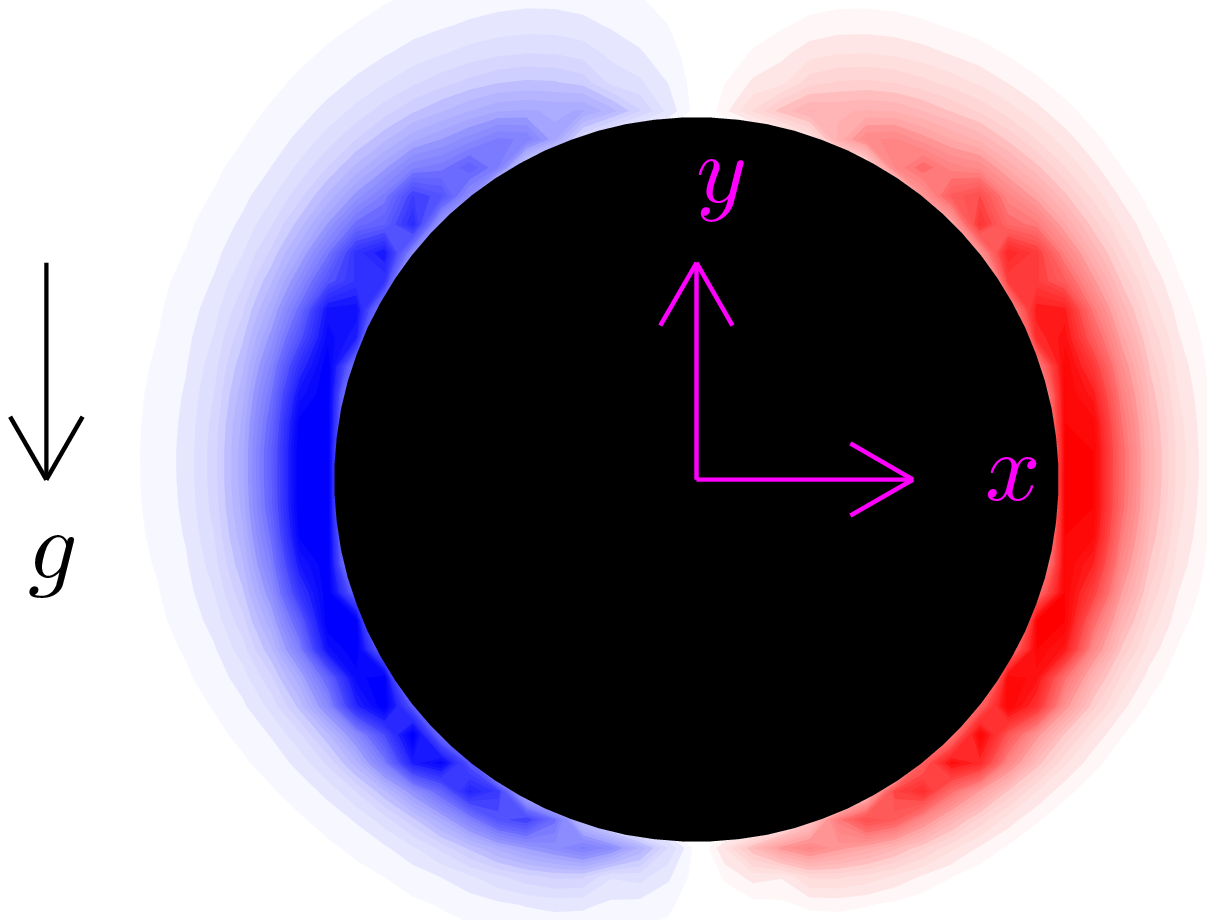}
		\caption{ }
		\label{fig_schematic}
	\end{subfigure}
	\begin{subfigure}[b]{0.34\textwidth}
		\includegraphics[width=\textwidth]{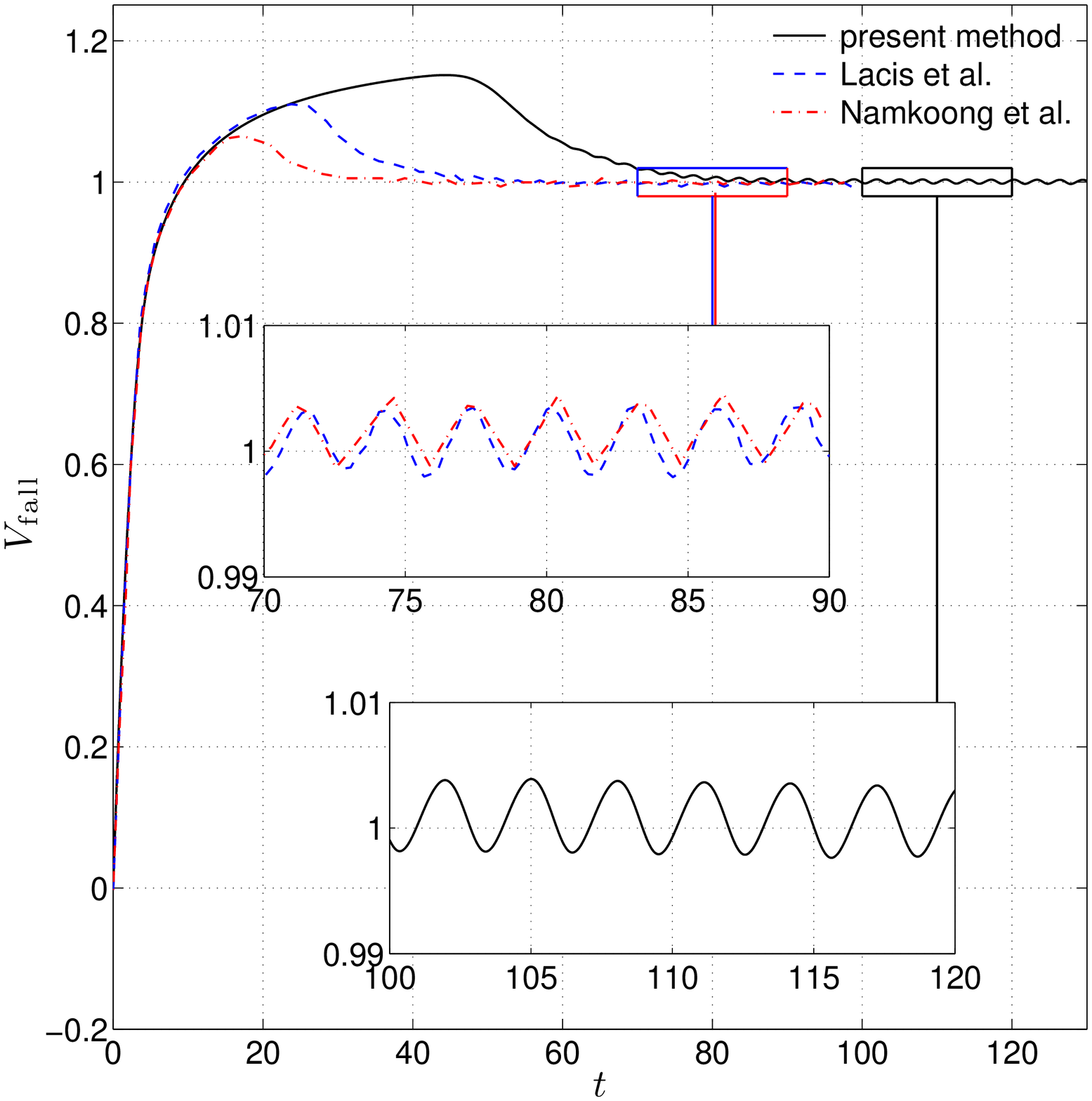}
		\caption{ }
		\label{fig_Vfall}
	\end{subfigure}
	\begin{subfigure}[b]{0.345\textwidth}
		\includegraphics[width=\textwidth]{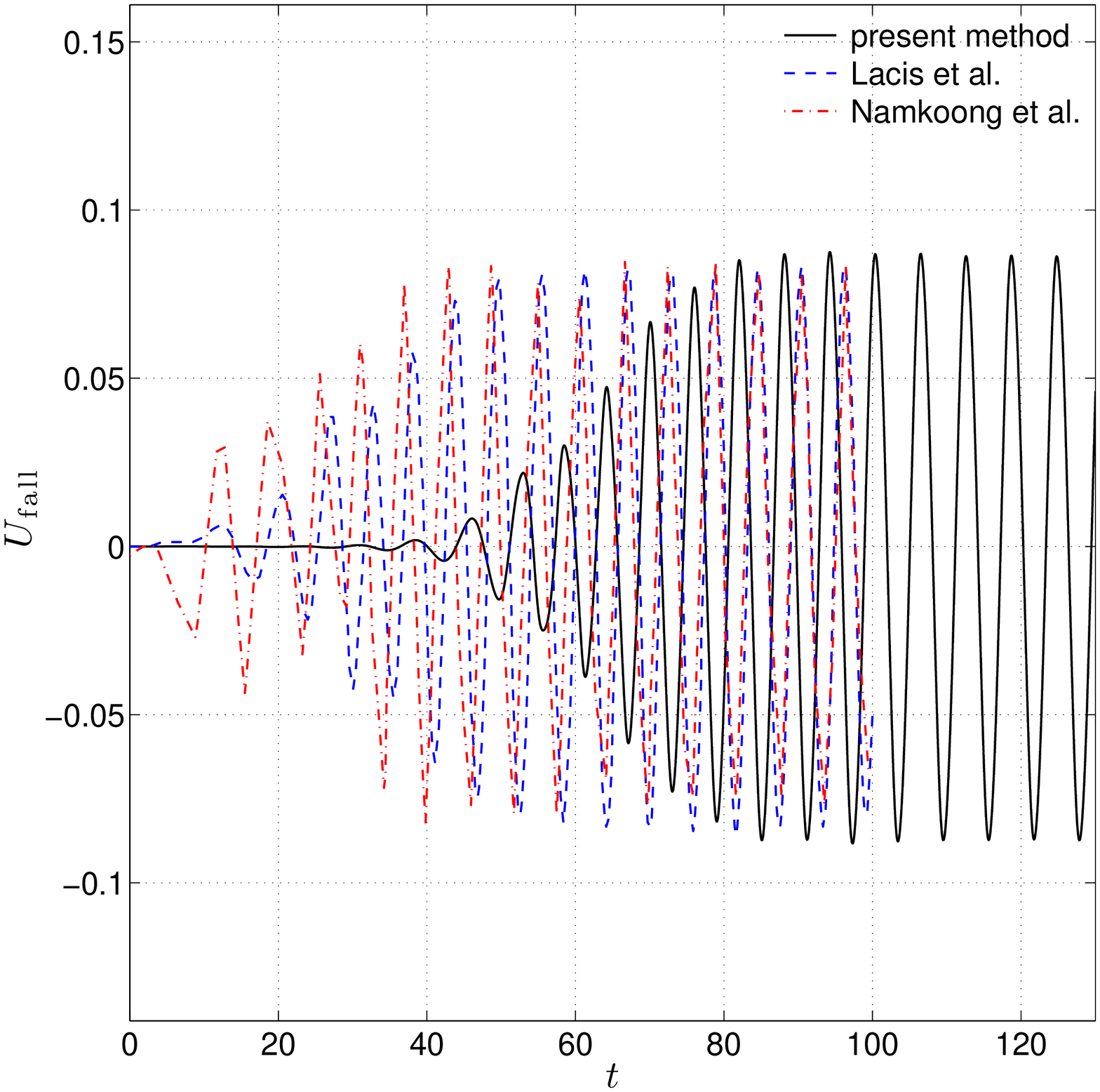}
		\caption{ }
		\label{fig_Ufall}
	\end{subfigure}
	\caption{(a) The schematic of a freely falling/rising cylinder and (b) the vertical and (c) transversal velocity of the freely falling cylinder with $\rho_s/\rho_f=1.01$ at $Re=156$ and $Ga=138$. Velocities from Namkoong {\it et al.} \cite{namkoong2008} and L{\=a}cis {\it et al.} \cite{lacis2016stable} are reported only up to $t=100$.}
	\label{fig_free_fall1}
\end{figure}

We consider a two-dimensional incompressible flow around a circular cylinder freely falling or rising under the gravitational pull, depending on the density of the cylinder, as shown in figure \ref{fig_schematic}. The characteristic length is the cylinder diameter $D$ and the characteristic velocity is the vertical terminal velocity $V_{\mathrm{term}}$. The origin of the body-fixed frame is placed at the center of the cylinder and the axes of the body-fixed frame are initially aligned in the horizontal and vertical directions. In order to validate our method, we compare the numerical solution with the works by Namkoong {\it et al.}\cite{namkoong2008} and L{\=a}cis {\it et al.}\cite{lacis2016stable}. Namkoong {\it et al.}\cite{namkoong2008} used a finite element method with implicit coupling and adaptive body-fitted mesh to simulate the flow in an infinite fluid, and the resolution in the wake was refined. L{\=a}cis {\it et al.}\cite{lacis2016stable} used an immersed boundary projection method in a finite domain of size $10D\times100D$ (with its origin at the center of the domain) with a time-lagged interpolation and an added-mass correction. A solid-to-fluid density ratio $\rho_s/\rho_f = 1.01$ for a freely falling cylinder and  $\rho_s/\rho_f = 0.99$ for a freely rising cylinder, a Reynolds number $Re=V_{\mathrm{term}}D/\nu_f=156$, and a Galilei number $Ga = \sqrt{|\rho_s/\rho_f-1|gD^3}/\nu_f=138$ were selected, where $g$ is the gravitational acceleration.

\begin{figure}[h!]
	\centering
	\begin{subfigure}[b]{0.46\textwidth}
		\includegraphics[width=0.4\textwidth]{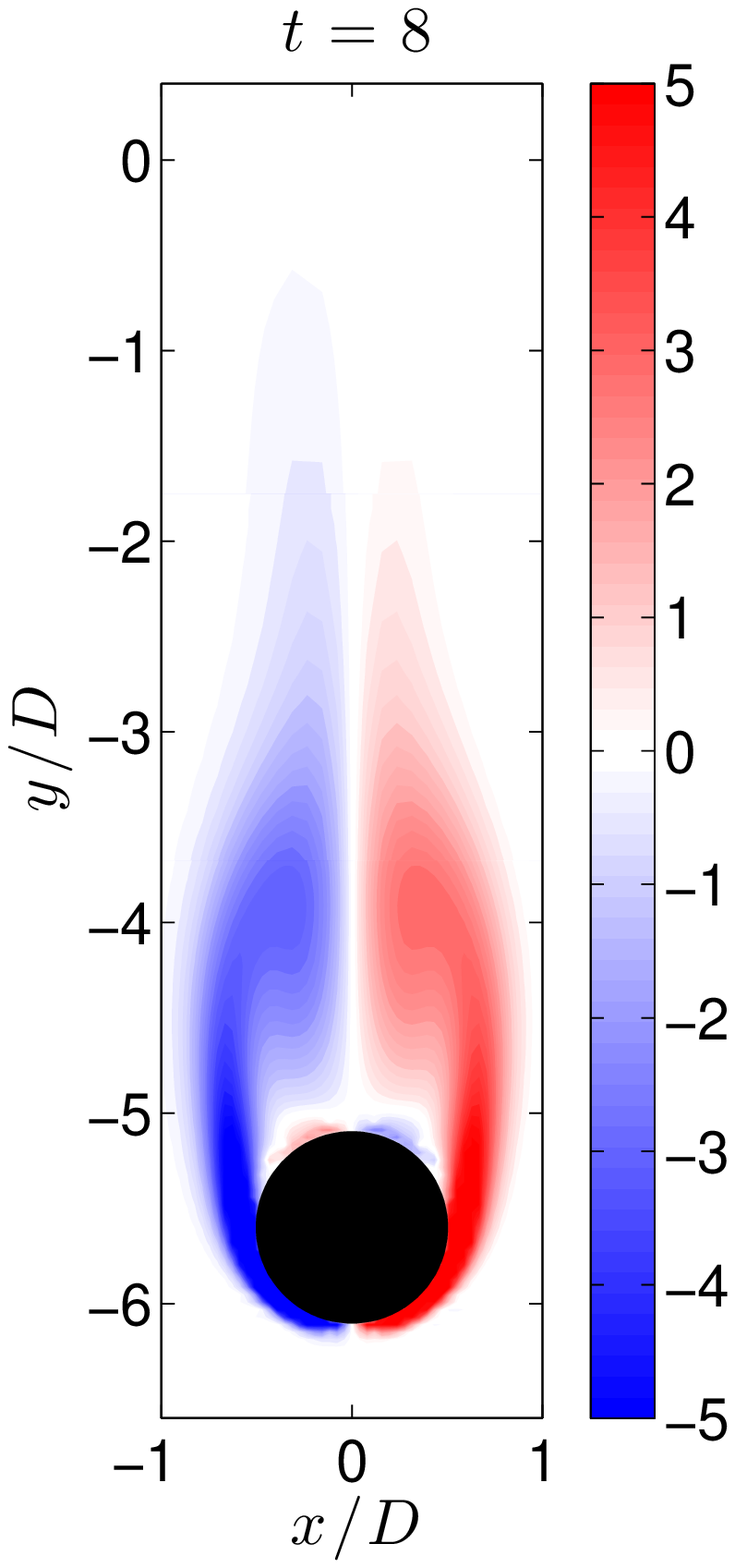}
		\includegraphics[width=0.52\textwidth]{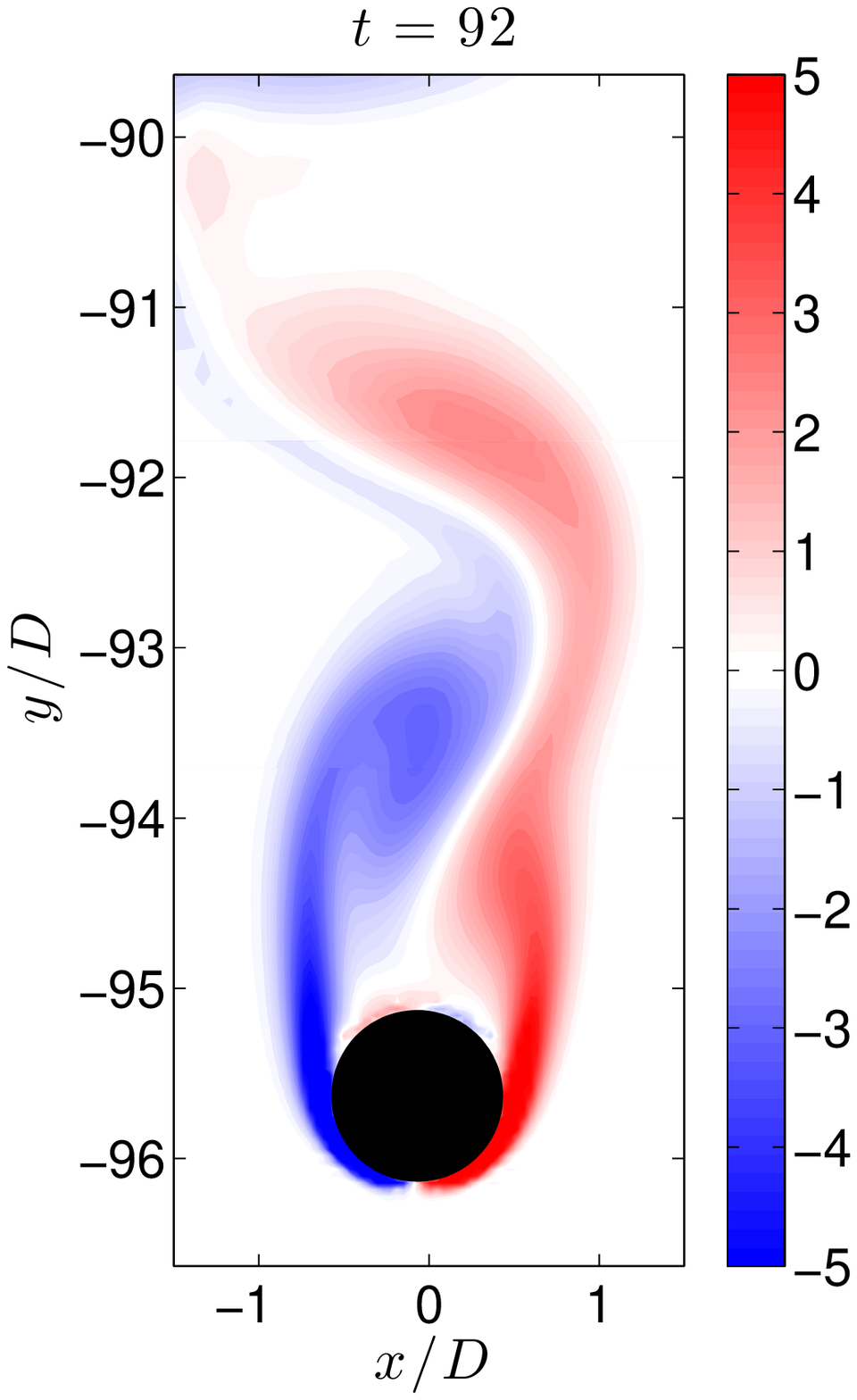}
		\caption{$\rho_s/\rho_f = 1.01$}
		\label{fig_fall}
	\end{subfigure}
	\begin{subfigure}[b]{0.445\textwidth}
		\includegraphics[width=0.395\textwidth]{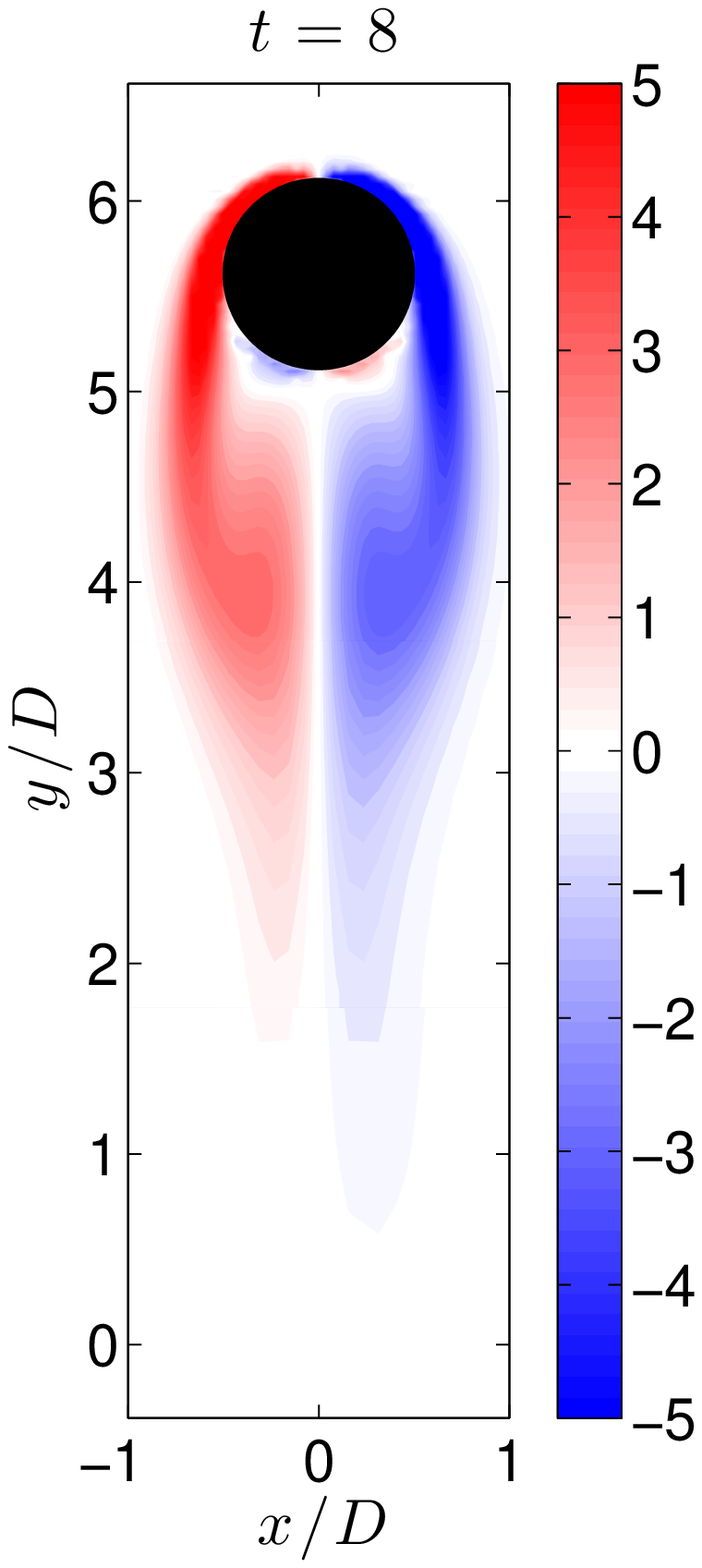}
		\includegraphics[width=0.52\textwidth]{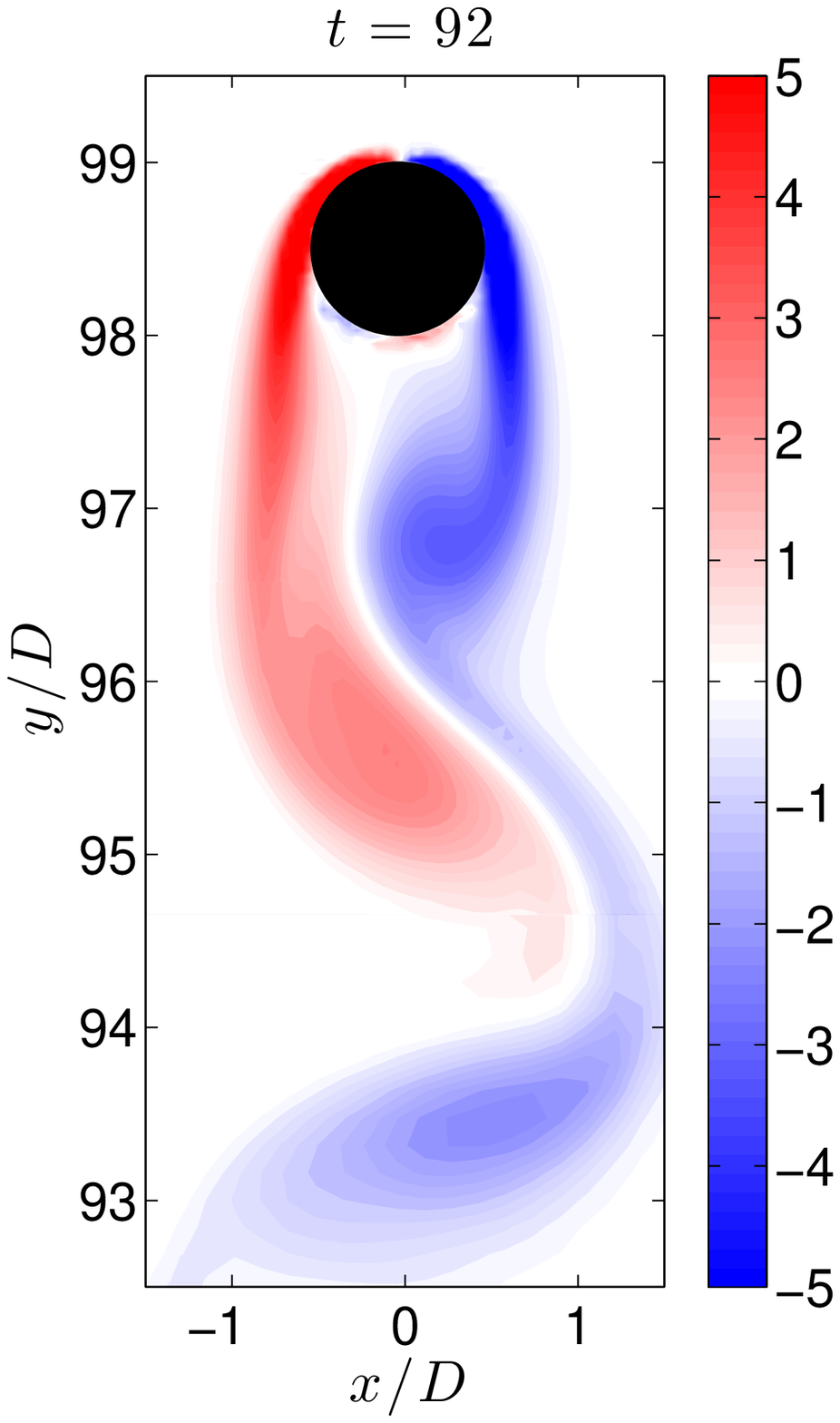}
		\caption{$\rho_s/\rho_f = 0.99$}
		\label{fig_raise}
	\end{subfigure}
	\caption{The vorticity fields of (a) a freely falling cylinder ($\rho_s/\rho_f = 1.01$) and (b) a freely rising cylinder ($\rho_s/\rho_f = 0.99$) at $t = 8$ and 92, $Re=156$, and $Ga=138$.}
	\label{fig_free_fall2}
\end{figure}

The multi-domain setting in our simulation is of the first domain size $4D\times4D$ and the number of domains $N_g = 6$. The finest grid spacing is $\Delta x = 0.04 D$ and the CFL number $V_{\mathrm{term}}\Delta t/\Delta x$ is set to be less than 0.4. Figure \ref{fig_Vfall} and figure \ref{fig_Ufall} show, respectively, the comparisons of the vertical and transversal velocity of the freely falling cylinder using the present method with that of reference \cite{namkoong2008} and \cite{lacis2016stable}. Figure \ref{fig_free_fall2} shows the vorticity fields around the freely falling and rising cylinders at $t = 8$ and 92. At an early time $t=8$, the cylinder wake is symmetric for both freely falling and rising cylinders. At a late time $t=92$, vortex shedding occurs in the wake for both cases. We can see from figure \ref{fig_Vfall} and figure \ref{fig_Ufall} that the vertical velocity agrees well in early development and later stationary oscillation, and that the transversal velocity reaches the same stationary oscillation at a later time. During the transient regime, the wake instability develops, breaks the symmetry of cylinder wake, and results in stationary vortex shedding (figure \ref{fig_fall}). The difference in the onset of wake instability observed in figure \ref{fig_Vfall} is due to the different rate at which numerical error accumulates. 

In Table \ref{tab_free_fall}, we compare the Strouhal number, the drag coefficient, and the amplitude of lift coefficient with L{\=a}cis {\it et al.}\cite{lacis2016stable} and Namkoong {\it et al.}\cite{namkoong2008}. The Strouhal number is defined as $St = f_L D/V_{\mathrm{term}}$, where $f_L$ is frequency of oscillations in lift force due to vortex shedding. The lift and drag coefficients are scaled by $\rho_fV_{\mathrm{term}}^2D/2$. Again, satisfactory agreements are observed, especially with the results of Namkoong {\it et al.}\cite{namkoong2008}, which are also simulated in an infinite fluid domain. We note that in both \cite{namkoong2008} and \cite{lacis2016stable} there is reported a slightly higher Strouhal number for a freely rising cylinder than a freely falling cylinder, while the Strouhal numbers for freely falling and rising cylinders are the same using the present method.

\begin{table}[h!]
	\caption{Flow characteristics of freely falling and rising cylinders with density ratios $\rho_s/\rho_f = 1.01$ and 0.99, respectively, at $Re=156$ and $Ga=138$. Values of Strouhal number, drag coefficient, and amplitude of lift coefficient are compared. We note that only the Strouhal number is reported for the freely rising cylinder in Namkoong {\it et al.}\cite{namkoong2008}.}
	\centering
	\begin{tabular}{@{}lllllllll@{}}
	\\ \hline
  & & \multicolumn{3}{l}{Falling} & & \multicolumn{3}{l}{Raising} \\ \cline{3-5}\cline{7-9}
	& & $St$  & $C_D$ & $\max|C_L|$  & & $St$  & $C_D$ & $\max|C_L|$ \\ \hline
	Present method					& & 0.1640	& 1.23 &	0.12 & & 0.1640 & 1.23 & 0.12\\
	L{\=a}cis {\it et al.}\cite{lacis2016stable} 	& & 0.17185	 & 1.29 &	0.14 & & 0.17188 & 1.29 & 0.14\\
	Namkoong {\it et al.}\cite{namkoong2008} & & 0.16840	& 1.23 &	0.15 & & 0.16870 & -- & --\\ \hline
	\end{tabular}
	\label{tab_free_fall}
\end{table}

\section{Conclusion}\label{sec_conclusion}
We presented an immersed boundary projection method that solves the flow-structure interaction problems involving rigid body kinematics. The method retains mathematical and computational simplicity by using the null-space approach and numerical stability for wide ranges of solid-to-fluid mass ratios by implementing the implicit coupling to rigid body dynamics. The effect of the fictitious fluid flow inside rigid bodies has also been taken into account to obtain accurate rigid body dynamics. The method further ensures the accuracy of surface stresses in space by applying an accurate stress filter, and in time by fixing the frame of reference on the target. We derive the developed method for general 3D rigid bodies and solve it efficiently by the block-LU decomposition. The method is validated for 2D flow around a freely falling or rising cylindrical rigid body and a neutrally buoyant cylinder migrating in a planar Couette flow. 

\section*{Acknowledgements}
This research was supported by a grant from the Ministry of Science and Technology, Taiwan (Grant No. MOST 108-2636-E-002-010).

\section*{Appendix A. Alternative form of the incompressible Navier-Stokes equations in a non-inertial frame of reference}\label{sec_appendixA}
	We start with the dimensionless incompressible Navier-Stokes equations and the dimensionless continuity equation in a non-inertial frame of reference that translates with velocity $\bm{U}(t)$ and rotates with angular velocity $\bm{\Omega}(t)$ about a center of rotation
	\begin{align}
		\nonumber
		\frac{\partial \bm{u}_n}{\partial t} + \left(\bm{u}_n\cdot\nabla\right)\bm{u}_n =& -\nabla p +\frac{1}{Re}\nabla^2\bm{u}_n + \bm{a} -\frac{d\bm{U}}{dt} - \bm{\Omega}\times\bm{U}\\
		&-\frac{d\bm{\Omega}}{dt}\times\bm{r} - 2\ \bm{\Omega}\times\bm{u}_n -\bm{\Omega}\times\left(\bm{\Omega}\times\bm{r}\right)\ ,\label{eq:non-inertial}\\
		\nabla\cdot\bm{u}_n =&\ 0\ ,\label{eq:non-inertial_con}
	\end{align}
	With the add of the facts that $\bm{a} = \nabla\left(\bm{a}\cdot\bm{r}\right)$, $\frac{d\bm{\Omega}}{dt}\times\bm{r} = \frac{d}{dt}\left(\bm{\Omega}\times\bm{r}\right) - \nabla[(\bm{\Omega}\times\frac{\partial \bm{r}}{\partial t})\cdot\bm{r}]$, $\bm{\Omega}\times\left(\bm{\Omega}\times\bm{r}\right) = -\nabla\left(\frac{1}{2}|\bm{\Omega}\times\bm{r}|^2\right)$ and $\bm{\Omega}\times\bm{U} = -\nabla\left[\left(\bm{\Omega}\times\bm{r}\right)\cdot\bm{U}\right]$, and the vector identity $ \left(\bm{u}_n\cdot\nabla\right)\bm{u}_n = \nabla\left(\frac{1}{2}\left|\bm{u}_n\right|^2\right) - \bm{u}_n\times\bm{\omega}_n$, (\ref{eq:non-inertial}) can be written as
	\begin{align}
		\frac{\partial \bm{u}_n}{\partial t}  = & -\nabla \Pi + \bm{u}_n\times\bm{\omega}_n +\frac{1}{Re}\nabla^2\bm{u}_n-\frac{d\bm{U}}{dt} -  \frac{d}{dt}\left(\bm{\Omega}\times\bm{r}\right) - 2\ \bm{\Omega}\times\bm{u}_n\ ,\label{eq:non-inertial2}
	\end{align}
	where
	\begin{align}
		\Pi = p + \frac{1}{2}\left|\bm{u}_n\right|^2 - \left(\bm{\Omega}\times\bm{r}\right)\cdot\bm{U} - \frac{1}{2}|\bm{\Omega}\times\bm{r}|^2 - \left(\bm{a}+\bm{\Omega}\times\frac{\partial\bm{r}}{\partial t}\right)\cdot\bm{r}\ .
	\end{align}
	
	We introduce the change of variables in velocity and vorticity
	\begin{align}
		\bm{u} &= \bm{u}_n + \bm{U} + \bm{\Omega}\times\bm{r} \equiv \bm{u}_n + \bm{U}_a\ ,\\
		\bm{\omega} &= \nabla\times\bm{u} = \bm{\omega_n} + 2\ \bm{\Omega}\ ,
	\end{align}
	where $\bm{U}_a = \bm{U} + \bm{\Omega}\times\bm{r}$ is the velocity of a fixed point in the non-inertial frame of reference relative to the inertial frame of reference, and $\bm{\omega}_n = \nabla\times\bm{u}_n$ is the fluid vorticity in the non-inertial frame. We can treat $\bm{u}$ and $\bm{\omega}$ as the fluid velocity and vorticity in the inertial frame, respectively. (\ref{eq:non-inertial_con}) and  (\ref{eq:non-inertial2}) can be written as
	\begin{align}
		\frac{\partial \bm{u}}{\partial t} = & -\nabla \Pi + \left(\bm{u} - \bm{U}_a\right)\times\bm{\omega} +\frac{1}{Re}\nabla^2\bm{u}\ ,\label{eq:non-inertial3}\\
		\nabla\cdot\bm{u} =&\ 0\ ,\label{eq:non-inertial_con2}
	\end{align}
	where
	\begin{align}
		\Pi = p + \frac{1}{2}\left|\bm{u} - \bm{U}_a\right|^2  - \frac{1}{2}\left|\bm{U}_a\right|^2  +  \frac{1}{2}\left|\bm{U}\right|^2 - \left(\bm{a}+\bm{\Omega}\times\frac{\partial\bm{r}}{\partial t}\right)\cdot\bm{r}\ .
	\end{align}
	(\ref{eq:non-inertial3}) and (\ref{eq:non-inertial_con2}) are not the standard non-inertial-frame form of equations, but are computationally convenient to implement because they renders the governing equations free from the body forces (eg. centrifugal forces, Coriolis forces, etc), and because the dependent variables decay at infinity.

\section*{Appendix B. Flow interacting with an impulsively rotated circular shell}
\label{sec_appendixB}

%\section*{References}

\bibliography{mybibfile}

\begin{thebibliography}{10}
\expandafter\ifx\csname url\endcsname\relax
  \def\url#1{\texttt{#1}}\fi
\expandafter\ifx\csname urlprefix\endcsname\relax\def\urlprefix{URL }\fi
\expandafter\ifx\csname href\endcsname\relax
  \def\href#1#2{#2} \def\path#1{#1}\fi

\bibitem{peskin1972flow}
C.~S. Peskin, Flow patterns around heart valves: a numerical method, J. Comput.
  Phys. 10~(2) (1972) 252--271.

\bibitem{taira2007projection}
K.~Taira, T.~Colonius, The immersed boundary method: a projection approach, J.
  Comput. Phys. 225~(2) (2007) 2118--2137.

\bibitem{colonius2008fast}
T.~Colonius, K.~Taira, A fast immersed boundary method using a nullspace
  approach and multi-domain far-field boundary conditions, Comput. Methods
  Appl. Mech. Eng. 197~(25) (2008) 2131--2146.

\bibitem{taira2009tip}
K.~Taira, T.~Colonius, Effect of tip vortices in low-reynolds-number poststall
  flow control, AIAA J. 47~(3) (2009) 749--756.

\bibitem{taira2009threedim}
K.~Taira, T.~Colonius, Three-dimensional flows around low-aspect-ratio
  flat-plate wings at low reynolds numbers, J. Fluid Mech. 623 (2009) 187--207.

\bibitem{taira2010lift}
K.~Taira, C.~W. Rowley, T.~Colonius, D.~R. Williams, Lift enhancement for
  low-aspect-ratio wings with periodic excitation, AIAA J. 48~(8) (2010)
  1785--1790.

\bibitem{chen2010}
K.~K. Chen, T.~Colonius, K.~Taira, The leading-edge vortex and quasi-steady
  vortex shedding on an accelerating plate, Phys. Fluids 22~(3) (2010) 033601.

\bibitem{choi2015}
J.~Choi, T.~Colonius, D.~R. Williams, Surging and plunging oscillations of an
  airfoil at low reynolds number, J. Fluid Mech. 763 (2015) 237--253.

\bibitem{li2012contact}
X.~B. Li, M.~L. Hunt, T.~Colonius, A contact model for normal immersed
  collisions between a particle and a wall, J. Fluid Mech. 691 (2012) 123--145.

\bibitem{tsai2016}
H.-C. Tsai, T.~Colonius, Coriolis effect on dynamic stall in a vertical axis
  wind turbine, AIAA J. 54~(1) (2016) 216--226.

\bibitem{uhlmann2005}
M.~Uhlmann, An immersed boundary method with direct forcing for the simulation
  of particulate flows, J. Comput. Phys. 209~(2) (2005) 448--476.

\bibitem{borazjani2008}
I.~Borazjani, L.~Ge, F.~Sotiropoulos, Curvilinear immersed boundary method for
  simulating fluid structure interaction with complex 3d rigid bodies, J.
  Comput. Phys. 227~(16) (2008) 7587--7620.

\bibitem{eldredge2008}
J.~D. Eldredge, Dynamically coupled fluid-body interactions in vorticity-based
  numerical simulations, J. Comput. Phys. 227~(21) (2008) 9170--9194.

\bibitem{kempe2012}
T.~Kempe, J.~Fr{\"o}hlich, An improved immersed boundary method with direct
  forcing for the simulation of particle laden flows, J. Comput. Phys. 231~(9)
  (2012) 3663--3684.

\bibitem{breugem2012}
W.-P. Breugem, A second-order accurate immersed boundary method for fully
  resolved simulations of particle-laden flows, J. Comput. Phys. 231~(13)
  (2012) 4469--4498.

\bibitem{yang2015}
J.~Yang, F.~Stern, A non-iterative direct forcing immersed boundary method for
  strongly-coupled fluid-solid interactions, J. Comput. Phys. 295 (2015)
  779--804.

\bibitem{wang2015}
C.~Wang, J.~D. Eldredge, Strongly coupled dynamics of fluids and rigid-body
  systems with the immersed boundary projection method, J. Comput. Phys. 295
  (2015) 87--113.

\bibitem{lacis2016stable}
U.~L{\=a}cis, K.~Taira, S.~Bagheri, A stable fluid--structure-interaction
  solver for low-density rigid bodies using the immersed boundary projection
  method, J. Comput. Phys. 305 (2016) 300--318.

\bibitem{zheng2010}
X.~Zheng, Q.~Xue, R.~Mittal, S.~Beilamowicz, A coupled sharp-interface immersed
  boundary-finite-element method for flow-structure interaction with
  application to human phonation, J. Biomech. Eng. 132~(11) (2010) 111003.

\bibitem{borazjani2013fsi}
I.~Borazjani, Fluid-structure interaction, immersed boundary-finite element
  method simulations of bio-prosthetic heart valves, Comput. Methods Appl.
  Mech. Eng. 257 (2013) 103--116.

\bibitem{goza2017coupled}
A.~Goza, T.~Colonius, A strongly-coupled immersed-boundary formulation for thin
  elastic structures, J. Comput. Phys. 336 (2017) 401--411.

\bibitem{tosi2019}
L.~P. Tosi, T.~Colonius, Modeling and simulation of a fluttering cantilever in
  channel flow, J. Fluids Struct. 89 (2019) 174--190.

\bibitem{yang2009}
X.~Yang, X.~Zhang, Z.~Li, G.-W. He, A smoothing technique for discrete delta
  functions with application to immersed boundary method in moving boundary
  simulations, J. Comput. Phys. 228~(20) (2009) 7821--7836.

\bibitem{seo2011}
J.~H. Seo, R.~Mittal, A sharp-interface immersed boundary method with improved
  mass conservation and reduced spurious pressure oscillations, J. Comput.
  Phys. 230~(19) (2011) 7347--7363.

\bibitem{goza2016accurate}
A.~Goza, S.~Liska, B.~Morley, T.~Colonius, Accurate computation of surface
  stresses and forces with immersed boundary methods, J. Comput. Phys. 321
  (2016) 860--873.

\bibitem{kallemov2016}
B.~Kallemov, A.~Bhalla, B.~Griffith, A.~Donev, An immersed boundary method for
  rigid bodies, Commun. Appl. Math. Comput. Sci. 11~(1) (2016) 79--141.

\bibitem{roma1999}
A.~M. Roma, C.~S. Peskin, M.~J. Berger, An adaptive version of the immersed
  boundary method, J. Comput. Phys. 153~(2) (1999) 509--534.

\bibitem{perot1993}
J.~B. Perot, An analysis of the fractional step method, J. Comput. Phys.
  108~(1) (1993) 51--58.

\bibitem{feng1994}
J.~Feng, H.~Hu, D.~Joseph, Direct simulation of initial value problems for the
  motion of solid bodies in a newtonian fluid. part 2. couette and poiseuille
  flows, J. Fluid Mech. 277~(271) (1994) 271--301.

\bibitem{vasseur1976}
P.~Vasseur, R.~G. Cox, The lateral migration of a spherical particle in
  two-dimensional shear flows, J. Fluid Mech. 78 (1976) 385.

\bibitem{namkoong2008}
K.~Namkoong, J.~Y. Yoo, H.~G. Choi, Numerical analysis of two-dimensional
  motion of a freely falling circular cylinder in an infinite fluid, J. Fluid
  Mech. 604 (2008) 33--53.

\end{thebibliography}

\end{document}